\documentclass[%
 aip,
 cha,
 amsmath,amssymb,
reprint,%
]{revtex4-1}

\usepackage{graphicx}% Include figure files
\usepackage{dcolumn}% Align table columns on decimal point
\usepackage{bm}% bold math
\usepackage[utf8]{inputenc}
\usepackage[T1]{fontenc}
\usepackage{mathptmx}
\usepackage{etoolbox}

\usepackage{graphicx}
\usepackage{pgfplots}
\pgfplotsset{compat=newest}
\usepackage{subcaption}

\usepackage{physics}
\usepackage{bm}
\usepackage{xcolor}
\usepackage{hyperref}

\newcommand{\bx}{\mathbf{x}}

\makeatletter
\def\@email#1#2{%
 \endgroup
 \patchcmd{\titleblock@produce}
  {\frontmatter@RRAPformat}
  {\frontmatter@RRAPformat{\produce@RRAP{*#1\href{mailto:#2}{#2}}}\frontmatter@RRAPformat}
  {}{}
}%
\makeatother
\begin{document}

\preprint{AIP/123-QED}

\title[Torus breakdown and collective chaos in heterogeneously forced excitatory--inhibitory neural populations with cascaded synapses]{Torus breakdown and collective chaos in heterogeneously forced excitatory--inhibitory neural populations with cascaded synapses}
% Force line breaks with \\
\author{Gonzalo Contreras-Aso}
 \email{gcontrer@math.uc3m.es.}
\affiliation{Grupo Interdisciplinar de Sistemas Complejos (GISC), Departamento de Matemáticas, Universidad Carlos III de Madrid, 28911 Leganés, Madrid, Spain}
\author{F. Javier Marín-Rodríguez}%
\affiliation{ 
Departamento de Matemática Aplicada, Ciencia e Ingeniería de los Materiales y Tecnología Electrónica, Universidad Rey Juan Carlos, 28933 Móstoles, Madrid, Spain %\\This line break forced with \textbackslash\textbackslash
}%

\author{Gabriel B. Mindlin}
 %\homepage{http://www.Second.institution.edu/~Charlie.Author.}
\affiliation{%
Universidad de Buenos Aires, Departamento de Física, Ciudad Universitaria, 1428, Buenos Aires, Argentina%\\This line break forced% with \\
}%
\affiliation{CONICET - Universidad de Buenos Aires, Instituto de Física Interdisciplinaria y Aplicada (INFINA), Ciudad Universitaria, 1428, Buenos Aires,
Argentina}

\date{\today}% It is always \today, today,

\begin{abstract}
Exact mean-field reductions of spiking networks provide low-dimensional
descriptions of neural populations that are derived from the microscopic
dynamics, rather than postulated phenomenologically. Here we extend the
Montbri\'o--Paz\'o--Roxin framework to excitatory--inhibitory networks of
quadratic integrate-and-fire neurons coupled through a double synaptic
filter, in which every spike elicits a second-order (alpha-function)
synaptic response. The reduction remains exact and yields an
eight-dimensional system. A two-parameter bifurcation analysis in the mean
excitabilities of both populations reveals nine dynamical regimes, organized
by saddle-node, SNIC, Hopf and homoclinic bifurcations, a Bautin point and
folds of cycles, including bistability between quiescence and collective
oscillations, which we confirm with network simulations. We then
periodically force the inhibitory population close to the homoclinic
bifurcation, where the collective rhythm is slow and effectively
two-dimensional. Weak forcing already produces quasiperiodicity, Arnold
tongues and chaos, consistent with torus breakdown. Remarkably, forcing only
a fraction $p$ of the inhibitory neurons with amplitude $A$ yields nearly the
same dynamics, including the Lyapunov exponents, as forcing all of them with
amplitude $pA$. We show that this equivalence follows from an exact rewriting
of the reduced equations, with corrections of order $p(1-p)A^2$, and that
heterogeneity guarantees the contraction between forced and unforced neurons,
making it a generic property of heterogeneous mean-field populations.
\end{abstract}

\maketitle

\begin{quotation}
Brain rhythms emerge from the collective activity of large populations of
neurons, and they are constantly shaped by rhythmic inputs coming from other
brain regions or from external stimulation. Models that describe such
populations through a few collective variables are essential to understand
this interplay, and over the last decade it has become possible to derive
them exactly from networks of spiking neurons~\cite{montbrio2015macroscopic,luke2013complete}.
We use this approach for interacting excitatory and inhibitory populations
whose synapses respond to each spike with a smooth rise and decay, and we map
the collective behaviors they can display. Close to the point where the
collective rhythm vanishes into a quiescent state, a weak periodic input
suffices to make the activity chaotic. Strikingly, what matters is
essentially the product of the input strength and the fraction of neurons
receiving it: stimulating few neurons strongly or many neurons weakly has
nearly the same effect. We explain why, and argue that this holds for
heterogeneous populations well beyond our model, as long as the input is
weak and not concentrated on too few neurons.
\end{quotation}

% \textcolor{red}{NOTACIÓN: distinta para variables micro y macro, +- o ei, fijar convención para forzado ($\omega_f, \Omega_f$). \\
% Por otro lado: igual conviene omitir la parte de las ecuaciones macroscópicas con $z$, y pasar directos a las del qif}

%%%%%%%%%%%%%%%%%%%%%%%%%%%%%%%%%%%%%%%%
\section{\label{sec:intro}Introduction}

The collective dynamics of large neural populations sits at the heart of theoretical neuroscience, because the variables accessible to experiment (firing rates, local field potentials, electroencephalographic signals)  are macroscopic averages over thousands of heterogeneous, interacting neurons. Moreover, when neural activity serves a behavioral purpose, it must ultimately be transduced into motor commands driving a peripheral biomechanical system; a musculoskeletal apparatus whose inertial and elastic properties act as a further dimensionality reducer, selecting only a low-dimensional subspace of possible movements from the vast output space of the network. Bridging this macroscopic level of description to the biophysics of individual spiking neurons is therefore both a mathematical challenge and a prerequisite for understanding how collective neural dynamics gives rise to coordinated behavior.

The phenomenological Wilson-Cowan model\cite{wilson1972excitatory} addressed this need early on, proposing evolution equations for the average activity of coupled excitatory and inhibitory populations through sigmoidal gain functions. While remarkably successful at capturing qualitative neural dynamics — bistability, oscillations, and their associated bifurcations — this model is not derived from a microscopic description of spiking neurons, which limits its interpretive reach and its connection to biophysical parameters.

A principled alternative became available with the Ott-Antonsen (OA) ansatz\cite{ott2008low}, which provides an exact dimensional reduction for large populations of coupled phase oscillators with Lorentzian-distributed heterogeneity. Applied to the theta neuron (a canonical model for Type-I excitability, whose onset of spiking occurs through a saddle-node bifurcation on a limit cycle\cite{kopell1986subcellular}) the OA method yields exact closed equations for macroscopic order parameters directly from the microscopic dynamics. Luke, Barreto and So\cite{luke2013complete} carried out a complete classification of the macroscopic behavior of heterogeneous theta-neuron networks using this approach, identifying the full repertoire of attractors and their bifurcations in parameter space. In a parallel line of work, Laing\cite{laing2014derivation} extended the OA reduction of theta-neuron networks to spatially structured settings, deriving neural field equations that retain the exactness of the mean-field description while allowing for spatial variation in connectivity; subsequent work has shown the framework to be broadly applicable across gap-junction and conductance-based coupling schemes as well\cite{bick2020understanding}.

The equivalence between the theta neuron and the quadratic integrate-and-fire (QIF) model was exploited by Montbrió, Pazó and Roxin\cite{montbrio2015macroscopic} to derive exact firing-rate equations (FREs) for heterogeneous, all-to-all coupled networks in the thermodynamic limit. Their mean-field description, cast in terms of the firing rate and mean voltage of the population, captures synchronous and asynchronous states alike and goes substantially beyond prior firing-rate theories limited to asynchronous regimes. Crucially, when the entire population is driven by a common periodic current, this framework already predicts macroscopic chaos through the interaction of the forcing with the spiral dynamics around an unstable fixed point; an exact collective chaotic state persisting in the thermodynamic limit.

The extension to coupled excitatory-inhibitory networks was carried out by Roulet and Mindlin\cite{roulet2016average}, who derived mean-field equations for two interacting theta-neuron populations via the OA method and obtained analytical expressions for their spiking rates. Comparing the resulting bifurcation diagrams with those of the Wilson-Cowan model, they found broad agreement in qualitative structure (bistability, Hopf-born oscillations, and global bifurcations) alongside higher-dimensional chaotic attractors absent from the phenomenological theory. Collective chaos in neural populations had also been identified along other routes: Pazó and Montbrió\cite{pazo2016quasiperiodic} traced its origin to quasiperiodic partial synchronization in inhibitory networks with delay, while So, Luke and Barreto\cite{so2014networks} found macroscopic chaos and multistability in theta-neuron networks with time-varying excitability, and Olmi, Politi and Torcini\cite{olmi2010collective} demonstrated collective chaos in pulse-coupled networks more broadly. A common thread in all these works, however, is the treatment of synaptic interactions as either instantaneous or governed by a single first-order exponential filter — a significant departure from the recorded shape of postsynaptic currents, which rise over a finite time before decaying, following a profile well approximated by a product of a linear ramp and a decaying exponential, i.e. a second-order (double-cascade) synaptic kernel \cite{boari2019observable,wilson1999spikes}.

The consequences of replacing instantaneous synapses with finite-timescale kinetics in exact mean-field theories have begun to receive attention. Dumont and Gutkin\cite{dumont2019macroscopic} introduced synaptic dynamics into the QIF mean-field framework via a first-order filter, finding that it reshapes the phase-resetting properties of oscillating networks. Barrio et al.\cite{barrio2024synaptic,mayora2025dynamics} subsequently carried out a detailed bifurcation analysis comparing the instantaneous-synapse (MPR) and first-order-synapse (DG) models for excitatory-inhibitory networks, finding that finite synaptic time constants suppress autonomous chaos and give way to spike-adding bursting dynamics, linked to a rich three-dimensional bifurcation skeleton. A second-order (double-cascade) filter, which produces a postsynaptic profile rising as $s(t)=te^{-t/\tau}$ before decaying (closer to what is observed physiologically) has been studied in the excitatory-only context by Boari et al.\cite{boari2019observable}, but its effect on the dynamics of a full excitatory-inhibitory network, and in particular on the system's capacity to generate and sustain collective chaos, has not been explored.

A separate question concerns what happens when the network is not driven uniformly, but rather when only a fraction of the population receives an external periodic stimulus. This situation is neurobiologically natural: rhythmic inputs from thalamic nuclei, sensory streams, or other cortical areas typically reach only a subset of neurons in a given population\cite{cardin2009driving, sohal2009parvalbumin}, and the question of how collective dynamics responds to such partial, heterogeneous drive is largely open. In the excitatory-only setting, Uribarri and Mindlin\cite{uribarri2019resonant} identified resonant features in a periodically forced excitatory population, but the excitatory-inhibitory case under heterogeneous drive has not been analyzed. The OA ansatz is in principle well suited to this problem: a bimodal forcing distribution $\Gamma(A) = p\delta(A-A_f) + (1-p)\delta(A)$, distinguishing forced and unforced neurons, splits the order parameter cleanly into two decoupled subpopulation contributions, each obeying its own OA-reduced equation while sharing the same synaptic feedback. This yields an exact, closed mean-field even under heterogeneous forcing; a non-trivial extension of the reduction that, to our knowledge, has not been previously reported for excitatory-inhibitory networks.

In this work we bring these threads together. We study the mean-field dynamics of coupled excitatory-inhibitory theta-neuron populations in which postsynaptic currents are governed by a second-order cascade, and subject a fraction of the inhibitory population to a periodic external drive. Applying the OA reduction and the QIF change of variables, we derive an autonomous eight-dimensional mean-field for the unforced system and characterize its bifurcation structure through numerical continuation, identifying nine dynamical regions delimited by saddle-node, sub- and supercritical Hopf, fold-of-cycles and homoclinic bifurcations. We then introduce heterogeneous periodic forcing, extending the mean-field to a twelve-dimensional non-autonomous system. Operating near a homoclinic orbit, we demonstrate that the forced system undergoes torus breakdown leading to collective chaos, and we identify two distinct routes to that chaos: increasing the forcing amplitude $A_f$ at full population coverage, or increasing the fraction $p$ of forced neurons at fixed amplitude. Characterizing both routes through maximum Lyapunov exponents and stroboscopic maps, we find a striking similarity, and whose geometric underpinning we explore through the toroidal structure of the attractor. The double-cascade synaptic model provides the autonomous dynamical substrate for this picture: the eight-dimensional mean-field it generates harbors a homoclinic orbit near which the forced system is operated, and it is the proximity to this orbit that makes collective chaos under partial periodic drive accessible.

% \textcolor{purple}{dejaría claro que la pregunta que estructura el paper es (sino me equivoco): ¿la población de neuronas sigue la misma ruta al caos cuando al forzar incrementamos la intensidad a cuando se aumenta la fraccion de neuronas forzadas? quizás en el párrafo de las secciones que queda o en el párrafo inmediatamente anterior a este mismo.}

This manuscript is structured as follows. In Section II we present the models and asumptions we will be working with, both from a microscopic and a macroscopic point of view, connected via the Ott-Antonsen reduction. The macroscopic dynamics allow us to write analytical expressions for the fixed points of the system in the absence of external currents. In Section III we analyze in detail the different regimes present in the limit of fast synapses using both analytical expressions as well as numerical continuation, and we show the agreement between the microscopic and macroscopic versions. In Section IV we add a heterogeneous forcing to the system, affecting part of the population and thus modifying both the microscopic and the macroscopic dynamics. We then numerically investigate how does it affect certain oscillatory dynamics, focusing on the appearance of chaotic behavior from two different trajectories along parameter space, finding that they mirror each other. We end in Section V with some conclusions and outlook. 
%%%%%%%%%%%%%%%%%%%%%%%%%%%%%%%%%%%%%%%%

%%%%%%%%%%%%%%%%%%%%%%%%%%%%%%%%%%%%%%%%
\section{\label{sec:prelim}Preliminaries and homogeneous model}

% \subsection{}

In this section we introduce and motivate the model under consideration. It will have both a microscopic description, in terms of individual neuronal dynamics with their respective synaptic couplings \cite{kopell1986subcellular,dumont2019macroscopic}, as well as a macroscopic one, in terms of collective variables such as the firing rate and the mean membrane potential \cite{montbrio2015macroscopic, barrio2024synaptic}.

\subsection{Microscopic model}

The theta model \cite{kopell1986subcellular} is a minimal model displaying the characteristics of Type I excitability: the state of the neuron is represented as a phase variable, with its excitability subject to a saddle-node on invariant circle (SNIC) bifurcation. We choose to represent this variable as $\theta_n^a$, where the $n$ subscript identifies the specific neuron, while the $a\in\{e,i\}$ superscript indicates the population the neuron belongs to (exclusively excitatory or inhibitory, as per Dale's principle \cite{hoppensteadt1997weakly}).
\begin{align}\label{eq:thetamodel_excitinhib}
    \dv{\theta_n^a}{t}
    &=
    (1-\cos\theta_n^a)
    +
    (1+\cos\theta_n^a)\, I^a_n(t), \nonumber \\
    I^a_n(t)&=\eta_n^a + J^{a e}s^e - J^{a i}s^i + I_{n, \rm ext}^a(t).
\end{align}

Here, $I^a_n$ is the input contribution to neuron $n$, which includes an intrinsic excitability contribution ($\eta^a_n$), synaptic contributions from the excitatory ($J^{a e} s^e$) and inhibitory ($J^{a i} s^i$) populations, and possibly an external input current ($I^a_{n, \rm ext}$). For the time being we ignore this external current, and incorporate it in Section~\ref{sec:forced}. The activations $s^e,\, s^i$ comprise the collective response of the excitatory and inhibitory populations, respectively, and are defined later. Note that we have chosen the convention that $J^{ab}>0$, so the sign of each interaction is carried explicitly in $I^a_n$. This corresponds to current-based synaptic coupling \cite{montbrio2015macroscopic, roulet2016average}.

While the theta model might be natural for dynamicists, the change of variables $V=\tan(\theta/2)$ turns it into the quadratic integrate-and-fire (QIF) one,
\begin{equation}
   \dv{V_n^a}{t} = (V_n^a)^2 + I^a_n(t), 
\end{equation}
which is more natural from a biological point of view, since it describes the dynamics of the membrane potential \cite{izhikevich2007dynamical}. From the point of view of the phase model, a spike is identified with the crossing of $\theta=\pi$, i.e., $V\to+\infty$ followed by reinjection at $V=-\infty$; the full rotation of $\theta$ comprises the spike and the subsequent recovery.

The individual excitability parameters are sampled from population-specific distributions $g^a(\eta)$. We follow the standard literature in the choice of Lorentzian distributed parameters \cite{ott2008low,luke2013complete,montbrio2015macroscopic}, namely
\begin{equation}\label{eq:lorentzian}
    g^a(\eta)=\frac{\Delta^a}{\pi}\frac{1}{(\eta-\eta^a_0)^2 + (\Delta^a)^2}
\end{equation}
where the distribution is centered around $\eta_0^a$, with $\Delta^a$ the ``half-width at half-maximum'', quantifying the heterogeneity of the population around said center. This choice is motivated by the analytical treatment it allows via the residue theorem, although it has been shown that results hold approximately for other distributions\cite{ott2008low}. 
%\textcolor{red}{Note that the algebraic decay of the tails of the distribution is numerically challenging, although it can be dealt with\cite{roulet2016average} ESTO MEJOR COMENTARLO EN APŃDICE}. 

Throughout this work we consider the case where $\eta_0^e,\, \eta_0^i <0$, which corresponds to intrinsically excitable neurons. That is the hallmark of excitability: it is a sufficiently strong input what triggers the excursion through phase space. The neuron alone (constant $I_n^a=\eta_n^a<0$) remains in a quiescent (resting) stable state, corresponding to $V_n^a=-\sqrt{I_n^a}$. High enough input, be it synaptic or external, pushes the phase beyond the excitability threshold set by $\eta_n^a$. Nonetheless, the Lorentzian distributed parameters still leave a small fraction of the population self-oscillatory. %\textcolor{red}{alguna referencia biológica, hay neuronas que son así?}.

% \todo[inline]{Aquí va a ser necesario justificar bien la presencia de la cascada doble} 

\begin{figure}[h]
    \centering
    \includegraphics[width=1\linewidth]{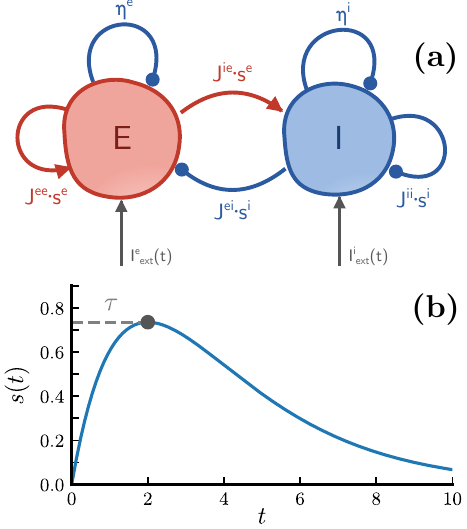}

    \caption{a) Diagrammatic representation of the excitatory and inhibitory populations, including their respective excitability parameters, synaptic couplings and external input currents. b) Synaptic activation function $s(t) = t e^{-t/\tau}$, with $\tau=2$.}
    \label{fig:WC_main}
\end{figure}

We consider activations in the form of secondary cascades, also known as second-order or alpha-function synapses \cite{rall1967distinguishing,wilson1999spikes}. These capture realistic post-synaptic rise and decay kinetics beyond instantaneous coupling. Dynamically, this cascade provides essential phase lag and additional degrees of freedom to sustain oscillations even without the presence of inhibitory neurons\cite{coombes2016next}. These second-order synapses are functions 
\begin{equation}\label{eq:activation}
s_n(t) \propto
\begin{cases}
(t-t_n) e^{-(t-t_n)/\tau}, & t>t_n, \\[0.4em]
0, & t \le t_n.
\end{cases}
\end{equation}
where $t_n$ denotes the instance at which the $n$th neuron received an input signal, and $\tau$ is the time difference between said reception and the maximum of the activity response, as shown in Figure~\ref{fig:WC_main}. This can be rewritten in a form amenable to subsequent analysis as the following system of first order differential equations,
\begin{equation}\label{eq:activationdifeqs}
    \begin{cases}
    \displaystyle \tau^a \dv{s_n^a}{t} = - s_n^a + w_n^a, \nonumber \\[0.1in]
    \displaystyle \tau^a \dv{w_n^a}{t} = - w_n^a + \delta(t-t_n), 
     \end{cases}
\end{equation}
where we already considered the excitatory and inhibitory populations, and an auxiliary variable $w_n^a$ is introduced. The $\tau^a$ parameters set a timescale for the synaptic dynamics, compared to the excitability scale (which can be thought of as being measured in units of the membrane time constant $\tau_m$). 

The collective response of each population is then encoded in the macroscopic variables appearing in equation \eqref{eq:thetamodel_excitinhib}, the average activities $s^e(t)=\langle s_n^e(t) \rangle$ and $ s^i(t) = \langle s_n^i(t)\rangle$, where $\langle\cdot \rangle$ denotes the average over the individual units.
% \begin{equation}\label{eq:macro_activations}
%     s^e=\frac{1}{n}\sum_{n} s_n^e, \quad s^i = \sum_{n} s_n^i,
% \end{equation}
% where the summation indices run over the respective populations.

% \missingfigure[figwidth=\linewidth]{Figura con dos poblaciones, y sus conexiones. Además la activación de cascada doble.}

\subsection{Macroscopic variables}

In 2008, an exact low-dimensional reduction for the synchronization of phase oscillators was put forward by Ott and Antonsen\cite{ott2008low, ott2009long}. They showed that, under the assumption of Lorentzian distributed intrinsic frequencies, the dynamics of an all-to-all population of phase oscillators in the thermodynamic limit was two-dimensional: there is a reduction of the dynamics to an invariant manifold which is attracting. The success of their approach led to several extensions in a variety of settings \cite{bick2020understanding}, including populations of theta and QIF neurons\cite{luke2013complete, montbrio2015macroscopic}. This was a substantive accomplishment, since up until that point macroscopic neuronal models were either phenomenological or derived under approximations, whereas the Ott-Antonsen reduction derives them exactly from microscopic dynamics.

The interested reader is referred to Appendix~\ref{appen:OA} for the derivation and the relevant literature\cite{luke2013complete, montbrio2015macroscopic, roulet2016average, boari2019observable}. It is possible to arrive at the following equations for the macroscopic order parameters $z^a(t)=\langle e^{i\theta_n^a(t)}\rangle$, with the novel addition of the double activation cascade:
\begin{align}
    \begin{cases}
        \displaystyle \dv{z^a}{t} = \displaystyle 
        - \dfrac{i}{2}(1-z^a)^{2} +\dfrac{1}{2}(1+z^a)^2[i(\eta_0^a+J^{ae} s^e - J^{ai} s^i) - \Delta^a],\\[0.1in]
        \displaystyle \tau^a \dv{s^a}{t} = - s^a + w^a,  \\
        \displaystyle \tau^a \dv{w^a}{t} = - w^a + \phi^a(t),
    \end{cases}
\end{align}
where the synaptic variables now include the accumulated effect of all microscopic synapses, i.e. $s^a= \sum_n s_n^a$ and likewise for $w^a$. We also introduced the spiking/firing rate $\phi^a(t)$, defined as
\begin{equation}
    \phi^a(t) = \frac{2}{\pi}
    \left[
    \frac{1+\Re(z^a)}{|1+z^a|^{2}} - \frac{1}{2}
    \right].
\end{equation}

% \todo[inline]{Aquí hablar brevement sobre la derivación del modelo macroscópico para $z$ hecha con OA por Luke, Barreto, So} 

Similarly to the situation with the theta and QIF neurons, the description of the collective state of the ensemble in terms of the order parameter may be more natural from a synchronization point of view, but it is less directly interpretable in physiological terms. Fortunately, one can apply a conformal transformation $(1-z)/(1+z) = \pi r - iv$, arriving to\cite{montbrio2015macroscopic} 

% \todo[inline]{Luego comentar que Montbrio, Pazo, Roxin lo rederivan a partir del QIF, y muestran que está relacionado mediante una transformación conforme, llegando a las ecs siguientes} 

\begin{equation}\label{eq:QIFmacroscopic}
    \begin{cases}
        \displaystyle \dv{r^a}{t} = \dfrac{\Delta^a}{\pi} + 2v^a r^a,  \\[0.1in]
        \displaystyle\dv{v^a}{t} = \eta_0^a + J^{ae} s^e - J^{ai} s^i  - \pi^2 (r^a)^2 + (v^a)^2,\\[0.1in]
    \displaystyle \tau^a \dv{s^a}{t} = - s^a + w^a,  \\[0.1in]
    \displaystyle \tau^a \dv{w^a}{t} = - w^a + r^a, 
     \end{cases}
\end{equation}
where $a\in\{e,i\}$. Rescaling time and the macroscopic variables with $\Delta^e$ reduces the number of parameters by one (Appendix~\ref{appen:adim}).

These equations represent the firing rate $r^a(t)=\phi^a(t)$ and the  mean membrane potential $v^a(t)$ (corresponding to the principal-value mean of $V_n^a$) of the ensemble of neurons of each kind, whose interplay provides the extra dimensionality, and its accompanying dynamical richness, which original activation models (such as the Wilson-Cowan one\cite{wilson1972excitatory}) lack. In fact, it has been shown that in the single cascade version, the limit of slow synapses formally recovers the Wilson-Cowan equations\cite{devalle2017firing}. The firing rates and mean membrane potentials are, in our work, subject to the double filter implemented by the secondary activation cascade. Note that the conformal or Möbius transformation maps $|z^a|<1$ to $r^a>0$, as it should due to the positivity of the firing rate.

This model is not only more biologically interpretable, but it is also more amenable to analytical treatment: fixed points satisfy the following closed form equations
\begin{widetext}
    \begin{align}\label{eq:fixedpointsQIFexcitinhib}
        v^a=-\frac{\Delta^a}{2\pi r^a}, \quad  r^a= s^a = w^a, \quad 
        \begin{cases}
            \pi^2(r^e)^4 - J^{ee}(r^e)^3 + J^{ei}(r^e)^2 (r^i) -\eta_0^e (r^e)^2 -\left(\dfrac{\Delta^e}{2\pi}\right)^2=0, \\[0.1in]
            \pi^2(r^i)^4 - J^{ie}(r^e) (r^i)^2 + J^{ii}(r^i)^3 -\eta_0^i (r^i)^2 -\left(\dfrac{\Delta^i}{2\pi}\right)^2=0,
        \end{cases}
    \end{align}
\end{widetext}
and the Jacobian $D_F(x)$ can also be analytically obtained. We do not show it here for conciseness, but it can be found in (see Appendix~\ref{appen:calc}).

From the point of view of the existence and stability of fixed points, there is a clear distinction between the parameters involved in the model. The specific choice of Lorentzian distributions (i.e. the values of $\eta_0^a,\,\Delta^a$) changes the location and existence of fixed points, but they do not enter the Jacobian explicitly. On the other hand, the time filters $\tau^a$ only enter the Jacobian, but not the fixed points themselves. This can also be seen from the Jacobian's determinant, which does not vanish for finite $\tau^a$. Regarding fixed points, these parameters can at most move the Hopf boundaries. %Due to the negativity of the coordinate $v^a$ at the fixed points, the trace of the Jacobian can be shown to always be negative, and thus there are no sustained oscillations regardless of the values of $\tau^a$. This is the case for the excitatory-only version, which despite the increase in dimensionality due to the double cascade, is still unable to display oscillations\cite{boari2019observable}.  

% \todo[inline]{Mencionar que ese es el model que vamos a utilizar. Mencionar que los puntos fijos, y el jacobian tienen expresiones analíticas manejables} 

% \todo[inline]{Mencionar que se pueden reducir los parámetros de estas ecuaciones mediante adimensionalización} 

%%%%%%%%%%%%%%%%%%%%%%%%%%%%%%%%%%%%%%%%

% \newpage
%%%%%%%%%%%%%%%%%%%%%%%%%%%%%%%%%%%%%%%%
\section{\label{sec:global}Global behavior of the unforced dynamics}

In this section we analyze the dynamical landscape which is obtained from the mean-field equations \eqref{eq:QIFmacroscopic}. Since a complete exploration of the space of parameters is unfeasible, we have chosen to explore the different behaviors which arise upon changing the average excitability of the model. In other words, we obtain a two-parameter bifurcation diagram in the $(\eta_0^e,\eta_0^i)$ plane. Regarding the remaining parameters, we choose them as
\begin{equation}\label{eq:macroparameters}
    \Delta^{e},\,\Delta^{i}=0.1,\quad \tau^{e},\,\tau^{i}=0.01,\quad J=\begin{pmatrix}
        J^{ee} & J^{ei}\\ J^{ie} & J^{ii}
    \end{pmatrix}=\begin{pmatrix}
        8 & 6\\ 6 & 4
    \end{pmatrix}.
\end{equation}

By the rescaling of Appendix~\ref{appen:adim}, the value of half-width at half-maximum of one of the two Lorentzian distributions, for instance $\Delta^e$, can be
fixed without loss of generality; we take $\Delta^e=0.1$ and, for simplicity,
$\Delta^i=\Delta^e$, so that the ratio $\Delta^i/\Delta^e$, which cannot be
scaled out, equals one.
% The choice of half-width at half-maximum of the Lorentzians $\Delta^{e},\,\Delta^{i}$ is completely arbitrary, since its definition can be absorbed into the others upon adimensionalization, see Appendix~\ref{appen:adim}.
The choice of time filters $\tau^e,\,\tau^i$ is motivated by the limit of fast synapses,% \textcolor{red}{hablar del límite y su inspiración biológica}
and, as we discussed in the previous section, it does not play a role in the existence and concrete value of the system's fixed points. Lastly, the coupling matrix $J$ is chosen so as to be able to sustain neural oscillations via the usual E-I feedback loop (known from the neurological literature as PING, \textit{pyramidal-interneuron network gamma}\cite{borgers2003synchronization}): the excitatory units require high self coupling to withstand their own intrinsic inhibition $(\eta_0^e<0)$ as well as inhibitory input from the other population. The inhibitory units engage with this dynamics as they receive an increasing excitatory synaptic input, with some delay due to the secondary cascade's time filtering, see Figure~\ref{fig:WC_main}. Their activation inhibits the entire system, bringing the system back to the initial state, which can start the cycle once more.

\begin{figure*}[t]
    \centering
    \includegraphics[width=\textwidth]{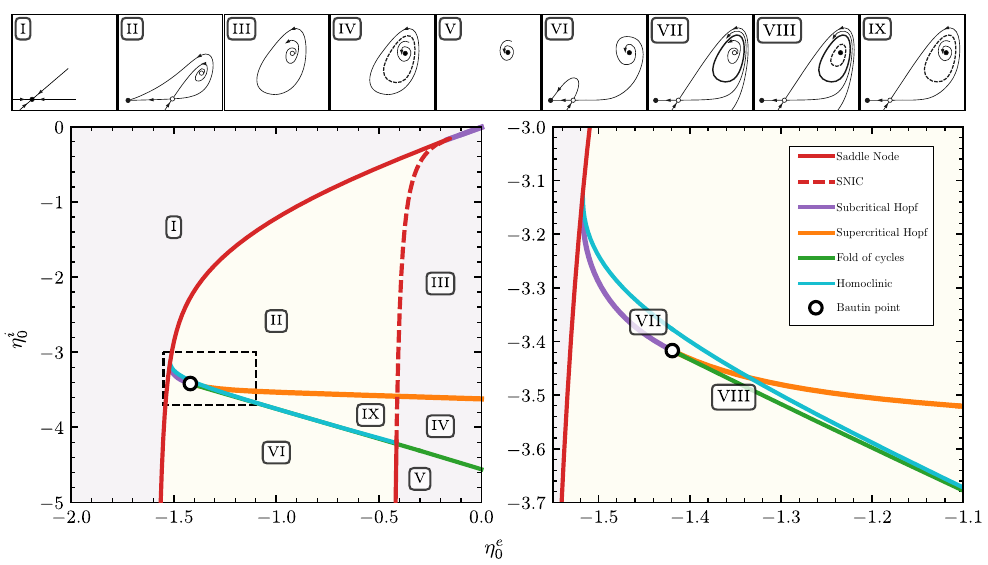}
    \caption{(Lower left) Two-parameter bifurcation diagram of \eqref{eq:QIFmacroscopic}, with parameters \eqref{eq:macroparameters}. (Lower right) zoom in the range $(-1.55,-1.1)\times(-3,-3.7)$, where regions VII and VIII can be seen more clearly. %Red lines indicate saddle-node, purple lines indicate supercritical Hopf, orange lines indicates subcritical Hopf, green lines indicate fold of cycles, and cyan lines indicate homoclinic bifurcations. The point is a ``Bautin point'', where the Hopf bifurcation changes behavior. 
    The region within the two saddle-node curves contains three equilibria, whereas the outside only features one equilibrium. 
    (Above) Illustrative, two-dimensional analogues of the phase portraits of each of the nine regions. 
    % \textcolor{red}{Tal vez mostrar el cambio a eigencoordinats en un apéndice?}
    }
    \label{fig:bifdiagram}
\end{figure*}

The phase diagram of Figure~\ref{fig:bifdiagram} is then constructed for values of $\eta_0^e\in(-2,0)$, $\eta_0^i\in(-5,0)$, using numerical continuation to obtain the lines associated to each bifurcation. This diagram has a clear qualitative resemblance to traditional heuristic firing rate models, such as  Wilson-Cowan (see, for instance Figure~2.13 of Hoppensteadt \& Izhikevich\cite{hoppensteadt1997weakly} or Figure~1 of Roulet \& Mindlin\cite{roulet2016average}),% \textcolor{red}{Me pregunto si sería interesante poner aquí una de esas Figuras o similares, referenciándola.},
with the signature cusp at small (in absolute value) mean excitability. The two saddle node bifurcations distinguish the number of equilibria in the system, with the region within them containing three fixed points, its outside only one. Another feature of that model which is present here is the existence of two Hopf bifurcations arising from the leftmost saddle node border, then going all the way to $\eta_0^e=0$. There are, however, notable differences with respect to this model, even at the qualitative level, which we discuss upon laying out the different regimes present in the model. 

\begin{figure}[h]
    \centering
    \includegraphics[width=0.95\linewidth]{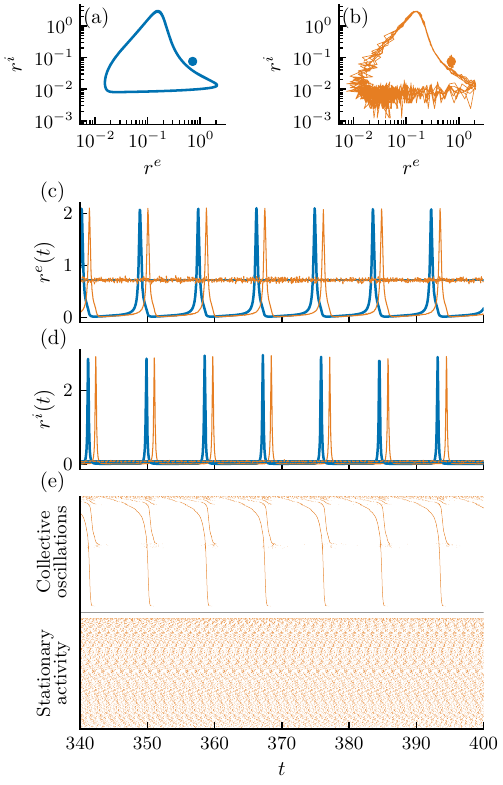}
\caption{
Bistability in region~IV. (a)~Macroscopic stable limit cycle (blue) and
stable equilibrium (blue dot) in the $(r^e,r^i)$ plane.
(b)~Same projection for the network: collective oscillations (orange)
and stationary activity (orange dot); the dot marks the
time-averaged stationary rates.
(c,d)~Excitatory and inhibitory firing rates for the two initial
conditions. Blue and orange curves: macroscopic and microscopic
oscillations; dark-blue solid and light-blue dashed curves:
macroscopic equilibrium and microscopic stationary activity.
(e)~Spike rasters of the same 1000 excitatory neurons in the
oscillatory (upper block) and stationary (lower block) states,
ordered by increasing excitability.
}
\label{fig:qif_bistability_regionIV}
\end{figure}

We find that the phase diagram distinguishes a total of 8 dynamical regimes\footnote{It is likely that the near-cusp region also contains more dynamical regimes, however it is clear from Figure~\ref{fig:bifdiagram} that it must be tiny. We do not resolve its structure, since it is unrelated to the latter part of our work.} labeled as I through to IX and illustrated\footnote{This is not computed from dynamics, for reasons related to the high dimensionality of the system. While great insight is obtained from the inspection of trajectories in the $(r^e,r^i)$ plane as in Figure~\ref{fig:qif_bistability_regionIV}, this projection makes visualizing the behavior worse: for instance, in said Figure we can see that the stable focus appears to be ``outside'' the limit cycle. One can fix this by using Hopf or phase-amplitude coordinates although in that case self-intersections appear in the limit cycle.} above the phase diagram of Figure~\ref{fig:bifdiagram}. These regions are separated by standard saddle node bifurcations colliding at the cusp, saddle node on invariant cycle (SNIC) bifurcations (which correspond to II$\leftrightarrow$III and IX$\leftrightarrow$IV), sub- and supercritical Hopf bifurcations colliding on a Bautin point (also known as a generalized Hopf\cite{Kuznetsov1998}), folds of cycles and homoclinic bifurcations. We also note the absence of the classic Bogdanov-Takens bifurcation at the point where the saddle node, Hopf and homoclinic curves meet; the fixed point undergoing the Hopf bifurcation is not one of the two involved in the saddle node bifurcation.  %\textcolor{red}{The fold of cycles one is not present in the heuristic firing rate models since it requires the added dimensionality to allow for the collision of periodic orbits.} 

Even though we have calculated the saddle-node bifurcations via numerical continuation, these can actually be written down analytically. This is done by solving for $\eta_0^\pm$ in the fixed point equations, obtaining 
\begin{equation}\label{eq:SNparametric}
    \begin{cases}
        \eta_0^e(\lambda) = \pi^2 \lambda^2 - J^{ee} \lambda + J^{ei} r^i(\lambda) - \dfrac{(\Delta^e)^2}{4\pi^2 \lambda^2}, \\[0.15in]
        \eta_0^i(\lambda) = \pi^2 \big(r^i(\lambda)\big)^2 - J^{ie} \lambda + J^{ii} r^i(\lambda) - \dfrac{(\Delta^i)^2}{4\pi^2 \big(r^i(\lambda)\big)^2},
    \end{cases}
\end{equation}
where $\lambda$ is the parameter of the curve and $r^{i}(\lambda)$ is the root of a quartic polynomial. The details are shown in Appendix~\ref{appen:calc}. One can check the absence of a $\tau^e,\tau^i$ dependence of these curves, as expected from the discussion of the previous section.

Note that, besides the added dimensionality and its new set of possible regimes, a notable difference with respect to the Wilson-Cowan model is the mechanism responsible for the destruction of the periodic orbits which are generated as we traverse the cusp's outer region, that is, the path I$\rightarrow$III$\rightarrow$IV$\rightarrow$V (and vice versa). In the standard Wilson-Cowan model, the transition between the stable fixed point states of regions I and V is mediated by two supercritical Hopf bifurcations\cite{hoppensteadt1997weakly}; oscillations are born and destroyed by the same mechanism. In this model, oscillations are born between I and III again by that same mechanism, however the transition between III and IV is supercritical: the unstable focus ``within'' the stable periodic orbit in region III is itself again undergoing a Hopf bifurcation, emitting an unstable orbit. In region IV a stable equilibrium coexists with a stable limit cycle: some initial conditions yield constant firing rates, others lead to neural oscillations. The transition between regions IV and V is then mediated by a fold of cycles; the stable and unstable orbits generated by the sequence of Hopf bifurcations meet, leaving just the stable fixed point alive. A byproduct of this is hysteresis: an oscillatory signal is lost at the fold of cycles, and is not recovered by inverting the process.

% \todo[inline]{Especificar los parámetros a usar, y motivarlos. Feedback loops}

% \todo[inline]{Hablar del límite de sinápsis rápidas  y su inspiración biológica} 

% \todo[inline]{Mencionar que vamos a construir el diagrama de bifurcaciones para el caso de etas principalmente negativos (aunque siendo sampleados de una Lorentziana serán más bien heterogéneos), como comentamos anteriormente}

% \todo[inline]{Mencionar cómo se ha obtenido mediante continuación el diagrama y hacer referncia a código disponible.} 

% \todo[inline]{Comentar las bifurcaciones y regímenes presentes. Comentar que en el pico es probable que haya una minúscula región con otras bifurcaciones como en WC tradicional} 

% \todo[inline]{Comentar que la autointersección en el plano (r+, r-) se puede eliminar mediante una transformación no lineal de coordenadas pero que no se visualiza bien ni así. Apéndice?.} 

% {\color{purple}
To illustrate the correspondence between the macroscopic dynamics
and the underlying neuronal network, we consider a representative point
in region~IV, $(\eta_0^e,\eta_0^i)=(-0.2,-4.0)$, where
Eq.~\eqref{eq:QIFmacroscopic} exhibits coexistence of a stable
equilibrium and a stable limit cycle. We compare these attractors with
direct simulations of Eq.~\eqref{eq:thetamodel_excitinhib} for
$N_e=N_i=2^{14}$ neurons, with independently sampled Lorentzian
excitabilities (the same realization in both runs) and two different
initial conditions, one for each attractor. Inhibitory inputs enter with
a negative sign.

Figure~\ref{fig:qif_bistability_regionIV} shows that the finite network
supports stationary and collective oscillatory states associated with
the two macroscopic attractors. Population rates are computed from all
$N_e=N_i=2^{14}$ neurons in time bins of width $0.05$, with integration
step $0.002$, and the interval $340\leq t\leq400$ is shown without
temporal realignment. The raster neurons in panel~(e) are randomly
selected from the excitatory population.
The stationary rates fluctuate around the macroscopic equilibrium,
while the collective oscillations trace a trajectory with a geometry
similar to the macroscopic limit cycle in the $(r^e,r^i)$ projection.
The apparent irregularity of the microscopic trajectory at low
firing rates is accentuated by the logarithmic axes, which emphasize
relative fluctuations of the finite-network rate estimates.
Panels~(c,d) reveal a gradual phase drift between the macroscopic and
microscopic oscillations, so the agreement in projected geometry does not
imply exact temporal coincidence. The rasters in panel~(e) distinguish
recurrent population bursts in the oscillatory state from dispersed
firing in the stationary state.
Together, these observations illustrate how the bistability of the
macroscopic equations is expressed in the collective activity of the
finite network.
% }

% \todo[inline]{Mencionar la baja dimensionalidad de la dinámica}

The model we considered thus contains, as witnessed by the phase diagram, a variety of dynamical regimes, featuring fixed points, oscillations, multistability, and hysteresis. However, all the encountered phenomenology is effectively low dimensional, and so are the bifurcations which can be found. All the bifurcations encountered, namely saddle-node, Hopf, Bautin, homoclinic, and fold-of-cycle bifurcations, are compatible with planar dynamics. In contrast, no bifurcations associated with intrinsically higher-dimensional oscillatory dynamics, such as torus or period-doubling bifurcations of periodic orbits, were detected in the parameter region explored.

In relation with the previous paragraph, the model does not display chaotic behavior, for this choice of parameters. There are two main arguments explaining this fact: first, even though the system is eight-dimensional, the limit of fast synapses ($\tau^e,\tau^i\ll 1$) which we are working under is effectively suppressing four out of those dimensions, namely, the equations are well approximated by the adiabatic elimination of the synaptic variables. This is still insufficient to explain the absence of chaos in the dynamics. For that it is important to observe that the coupling between populations in \eqref{eq:QIFmacroscopic} is linear: the populations only inform each other via the linear coupling weighted by the matrix $J$. This does not rule out chaotic behavior, but it does make it harder to be present. The case for this is stronger when we consider that this linear coupling is also weak, since the synaptic variables $s^a(t)$ are generally small. %, in accordance with the phase diagram of Figure~\ref{fig:bifdiagram}. Moreover, 

We leave for future work the exploration of the synaptic regime where the two timescales are comparable; the slow synaptic dynamics were studied (albeit with a single activation cascade) by Devalle, Roxin \& Montbrió\cite{devalle2017firing}, where they showed that the variables $r,v$ follow their fixed points adiabatically. In turn, in that limit they find for a single population
\begin{equation}
    r^* = \Phi(\eta_0 + Js), \quad \tau \dot s = -s + \Phi(\eta_0 + Js)
\end{equation}
with $\pi\sqrt{2}\,\Phi(I) = \sqrt{I+\sqrt{I^2 + \Delta^2}}$. This rendered the dynamics of the global activation $s$ formally equivalent to that of the standard Wilson-Cowan model \cite{wilson1972excitatory,hoppensteadt1997weakly}.

% \todo[inline]{Mencionar la falta de caos como consecuencia de los acoples rápidos y la linealidad entre poblaciones} 

% \textcolor{purple}{Querría remarcar que me centraría en mostrar el mapa, caracterizar las regiones (eso ya está) y sobre todo hacer zoom a la órbita periódica macro/micro que es próxima a la homoclina que es realmente lo que guía todos los resultados de forzado (section IV).}

%%%%%%%%%%%%%%%%%%%%%%%%%%%%%%%%%%%%%%%%

% \newpage
%%%%%%%%%%%%%%%%%%%%%%%%%%%%%%%%%%%%%%%%
\section{\label{sec:forced}Dynamics under heterogeneous forcing}

A qualitative difference of the phase diagram in Figure~\ref{fig:bifdiagram} with respect to the traditional firing rate ones, concerns the region around the Bautin point. This point is also present in these models\cite{hoppensteadt1997weakly}, however, in that case the phenomenology is different: the homoclinic orbit is responsible for the birth of an unstable orbit around one of the three fixed points, whereas in our case the homoclinic gives birth to a stable one. This is also due to having the lower Hopf branch inverted with respect to Wilson-Cowan, with the supercritical bifurcation reaching towards the right ($\eta_0^e=0$) border. 

We thus turn our attention to region VII, between the Bautin point and the homoclinic bifurcation. Up until now we have considered the unforced model \eqref{eq:QIFmacroscopic}, but it is natural to wonder what periodic forcing does to the system, particularly at low amplitudes. From a biological point of view, the macroscopic systems under consideration (ensembles of neurons) are rarely in isolation, instead, there are periodic inputs constantly exerted over the individual units\cite{buzsaki2002theta}. It should be emphasized that under such a forcing the system becomes non-autonomous: the phase diagram we showed before is a mere guide for our expectations, but we can expect it to be more faithful for the small amplitude regime we consider. 

% \todo[inline]{habría que hacer un repaso rápido a lo que la gente ha visto con Wilson-Cowan forzado}

Periodic forcing of exact next-generation neural mass models has been
studied before in related settings. Schmidt \emph{et al.}~\cite{schmidt2018network}
forced QIF networks that are multistable in the absence of input and showed
that the outcome is set by the forcing frequency, while Ceni \emph{et al.}~\cite{ceni2020cross} considered inhibitory populations with first-order (exponential) synapses, whose collective oscillations arise at a supercritical Hopf bifurcation. Here we consider a different situation: the forced state is a collective oscillation of an excitatory--inhibitory network lying close to a homoclinic bifurcation, so that the forcing may not only lock or destabilize the rhythm but also drive the system onto a coexisting quiescent state. Moreover, following
Ref.~\cite{uribarri2019resonant}, we allow the forcing to act on only a
fraction of the inhibitory population.

In order to modify the model to include this forcing, we first need to establish \textit{who} is receiving the periodic forcing, and how strong it is. This in fact divides the population in different subsets, one per specific forcing. It turns out that the mean-field description of the population's activity does allow for different subpopulation forcings, even from an analytical standpoint\cite{uribarri2019resonant}. We now review how this heterogeneity in the forcing can be included in the macroscopic description of the system.

For simplicity, let us consider a single population (excitatory or inhibitory) which is divided into just two subpopulations, a forced and an unforced one. The forcing is implemented through the external input of \eqref{eq:thetamodel_excitinhib}, namely
\begin{equation}
    I_{n, \rm ext}(t) =A_f \cos(\Omega_f t),
\end{equation}
where $n$ is the affected neuron, $A_f$ is the amplitude of the forcing, and $\Omega_f$ its frequency.%\footnote{In principle it could also be offset by a constant factor $\phi$ as $A_f \cos(\Omega_f t + \phi)$ but we choose to ignore such effect for simplicity.}
The amplitude $A_f$ takes a nonzero value for the forced units, zero otherwise. The amplitudes are drawn at random from a distribution, which we assume takes the form
\begin{equation}
    \Gamma(A) = p \delta(A-A_f) + (1-p) \delta(A).
\end{equation}

For $p\neq0,1$ the corresponding population is effectively divided into the forced and unforced sectors. The density function needs to be updated to include this heterogeneity as $f(\theta,\eta,A,t)$, thus satisfying
\begin{equation}
    \int^{2\pi}_0 f(\theta,\eta,A,t)\, \text{d}\theta = g(\eta) \Gamma(A).
\end{equation}

Note that the assignment to the forced subset is independent of excitability, which gives both subpopulations the same Lorentzian. The Ott-Antonsen procedure can then be carried out as usual\cite{luke2013complete,montbrio2015macroscopic} (see Appendix~\ref{appen:OA}), with the subtlety that the order parameter of the entire population is now composed of a weighted sum of the order parameters of each subpopulation, that is, 
\begin{align}
z(t)
&= \int_{0}^{2\pi}\int_{-\infty}^\infty\int_{-\infty}^{\infty}
    f(\theta,\eta,A,t)e^{i\theta}\, \text{d}\eta\, \text{d}A\,\text{d}\theta \nonumber
\\
&= p \, z_f(t) + (1-p)\, z_u(t),
\end{align}
where the subpopulation order parameters are averages over the respective units,
\begin{equation}
    z_f(t) = \left\langle e^{i\theta_n(t)}\right\rangle_{\rm forced}, \quad z_u(t) = \left\langle e^{i\theta_n(t)}\right\rangle_{\rm unforced}.
\end{equation}

An analogous calculation shows that the activity/firing rate is also split as a weighted sum of the individual population's ones %$\phi(t)=p\phi_f(t) + (1-p) \phi_u(t)$. 
\begin{align}
    \phi(t)
    &= \int_{-\infty}^{\infty}\int_{-\infty}^{\infty}
\left[ f(\theta,\eta,A,t)\frac{d\theta}{dt}(\theta,\eta,A,t) \right]\bigg\rvert_{\theta=\pi}
 \text{d}\eta \, \text{d}A \nonumber \\
    &=p\,\phi_f(t) + (1-p)\, \phi_u(t) = p\,r_f(t) + (1-p)\, r_u.
\end{align}

Note that this population division involves doubling the degrees of freedom corresponding to the $(r,v)$ variables. The synaptic variables, however, remain the same population-wide, and are therefore affected by the collective response (the order parameter) of the ensemble. The generalization to more than one forcings at once over different subpopulations is straightforward\cite{uribarri2019resonant}, for instance with three subpopulations, the calculation would be redone for $\Gamma(A)=p\,\delta(A-A_p) + q\,\delta(A-A_q) + (1-p-q)\,\delta(A)$.

We now turn to our original problem, thus incorporating the two original populations (excitatory and inhibitory), identified by their respective superscripts, and having the subscripts $f,u$ identify the forced and unforced subpopulations, respectively. The resulting system of equations, upon changing to the QIF variables, is
\begin{equation}\label{eq:QIFmacroscopicForced}
    \begin{cases}
        \displaystyle \dv{r_f^a}{t} = \dfrac{\Delta^a}{\pi} + 2v_f^a r_f^a,  \\[0.1in]
        \displaystyle \dv{v_f^a}{t} = \eta_0^a + A_f^a\cos(\Omega^a_f t) + J^{ae} s^e - J^{ai}s^i - \pi^2 (r_f^a)^2 + (v_f^a)^2, \\[0.1in]
        \displaystyle \dv{r_u^a}{t} = \dfrac{\Delta^a}{\pi} + 2v_u^a r_u^a,  \\[0.1in]
        \displaystyle\dv{v_u^a}{t} = \eta_0^a + J^{ae} s^e - J^{ai}s^i - \pi^2 (r_u^a)^2 + (v_u^a)^2,  \\[0.1in]
        \tau^a \dfrac{d s^a}{dt} = - s^a + w^a,  \\
        \tau^a \dfrac{d w^a}{dt}
         = - w^a + p^a\,r_f^a + (1-p^a)\, r_u^a.
    \end{cases}
\end{equation}

This is a system of four equations for the unforced subpopulations' firing rate and mean membrane potential, four for the analogous variables of the forced subpopulation, and four for the double synaptic cascades of each population, comprising a twelve-dimensional space where the dynamics takes place. A caveat of this approach is the fact that it assumes the thermodynamic limit for each of the four subpopulations (unforced excitatory, unforced inhibitory, forced excitatory, forced inhibitory). While this is mathematically plausible, it makes numerical simulations much more prone to finite-size effects. %\textcolor{red}{This is something we discuss.}

We can finally start analyzing the effect of the forcing on the ensemble of excitable systems. In the coming subsections we analyze two scenarios: the homogeneous scenario, that is, forcing an entire population with different amplitudes and frequencies; and the heterogeneous scenario, where we fix the amplitude and frequency of the forcing and instead vary the fraction of the population which is affected by said external input.

\subsection{Varying forcing amplitude over the inhibitory population}

We now consider the case with $p^e=0, \, p^i=1$. It is well known that forcing an oscillatory system can induce resonances, locking and chaos when the forcing frequency is close to multiples of the natural frequency of the unforced system. In our case, we want to understand what happens with a small-amplitude oscillation, such as the ones in region VII. This is particularly interesting due to the presence of the homoclinic bifurcation nearby in parameter space. 

The interplay between forcing and homoclinic orbits is related to chaos via homoclinic tangles, via Melnikov's theory\footnote{Melnikov's theory is planar, however, our system is effectively two-dimensional near the orbit, as argued in the main text.}: at the homoclinic parameter value, an arbitrarily small periodic forcing produces transverse homoclinic intersections\cite{wiggins2003introduction}. Based on the phase diagram of Figure~\ref{fig:bifdiagram}, we note however that forcing the system precisely at the homoclinic parameter values between regions II and VII will destroy the oscillations when the forcing has pushed the system into region II (which has a single stable fixed point unrelated to the limit cycle). It is for that reason that we consider parameter values slightly off the homoclinic, therefore expecting chaos to emerge when the amplitude of the forcing is comparable to the distance from the homoclinic.

%\textcolor{red}{revisar todo esto. También, mirar la proporción de la amplitud del forzado respecto a la amplitud del ciclo.}

For $(\eta_0^e, \eta_0^i) = (-1.419,-3.38)$, before activating the forcing we find a stable, low activity fixed point $\bx_1^*$ and two saddles $\bx_2^*,\bx_3^*$ corresponding to intermediate and high activity, respectively. These fixed points have $(r^e,\, r^i)$ coordinates $(0.014, 0.009)$, $(0.29, 0.012)$ and $(0.49, 0.022)$ respectively, with the rest of the coordinates following directly from them via \eqref{eq:fixedpointsQIFexcitinhib}.
% \begin{equation}
%     \bx_1^* = \begin{pmatrix}
%         0.014\\ -1.168\\  0.014\\  0.014\\  0.009\\ -1.826\\  0.009\\  0.009
%     \end{pmatrix},\, 
%     \bx_2^* = \begin{pmatrix}
%         0.29\\  -0.055\\  0.29\\   0.29\\   0.012\\ -1.3\\    0.012\\  0.012
%     \end{pmatrix}, \, 
%     \bx_3^* = \begin{pmatrix}
%         0.49\\  -0.032\\  0.49\\   0.49\\   0.022\\ -0.729\\  0.022\\  0.022
%     \end{pmatrix},
% \end{equation}
%where the coordinates are ordered as $(r^e,v^e,s^e, w^e, r^i,v^i,s^i, w^i)$. 

The saddle at $\bx_2^*$ is the one whose homoclinic connection is broken in the homoclinic bifurcation, with the one at $\bx_3^*$ losing stability at the supercritical Hopf bifurcation from below, giving rise to the stable limit cycle that surrounds it. The limit cycle has, approximately, an angular frequency $\Omega\approx 0.6904$ in units of the excitability dynamics $\tau_m^{-1}$. This corresponds, for $\tau_m\sim10-20$ms, to forcing frequencies of $f=\Omega/(2\pi \tau_m)\sim12-24$Hz. 

Besides the trivial unit multiplier, the Floquet spectrum of this cycle
contains a single multiplier of order one, $\approx 0.53$, while the remaining
six have modulus below $10^{-7}$. The cycle therefore lies on a strongly
attracting two-dimensional invariant manifold. This is consistent with the
spectrum of the saddle $\bx_2^*$, whose leading eigenvalues are real
($\approx 0.98$ and $\approx -1.19$), so that the cycle born at the homoclinic bifurcation is stable~\cite{Kuznetsov1998}.

\begin{figure}[h]
    \centering
    \includegraphics[width=0.95\linewidth]{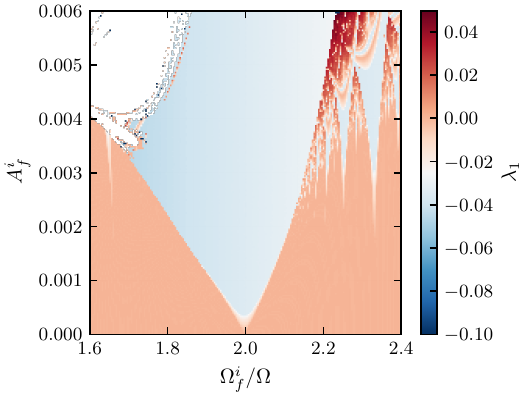}
    \caption{Maximum Lyapunov Exponent associated to each pair $(\Omega_f^i/\Omega, A_f^i)\in(1.6,2.4 )\times(0,0.006)$. The upper-left corner in white corresponds to the solution having converged to $\bx_1^*$.% Red lines mark the borders $\lambda_1= 2$ of chaotic regions.
    }
    \label{fig:forcing_scan}
\end{figure}

Forcing the entire inhibitory population around $2\Omega$ (thus targetting the 2:1 tongue), for values of the amplitude $A_f^i$ in the range $(0,0.006)$, we compute the Maximum Lyapunov Exponent (MLE) after a transient, obtaining the diagram in Figure~\ref{fig:forcing_scan}. This diagram presents a variety of behaviors: Arnold tongues, where the MLE of the non-autonomous system is negative and thus trajectories tend to attract each other; regions of quasiperiodicity with zero MLE; and chaotic regions, with a positive although small MLE.

% \todo[inline]{Comentar que nos situamos cerca de la homoclina, ya que es una región interesante por la competición entre un forzado y la homoclina, a través de Melnikov.}

% \todo[inline]{Identificar la órbita estable, caracterizar su periodo y aplicar el forzado con frecuencias similares y amplitud creciente}

This inhibitory forcing can be interpreted, on the level of parameter space, an oscillation of the system alternating towards and away from the homoclinic bifurcation. The basin of attraction of the limit cycle is quite narrow: for a high enough amplitude, this forcing destroys the oscillatory behavior of the systems, since beyond the homoclinic any trajectory flows to the stable fixed point. This is a point we already discussed in relation to Melnikov's theory. Notice that avoiding this requires using small amplitudes, as in Figure~\ref{fig:forcing_scan}, and despite this we find chaos unfolding for some of these forcings.

% \todo[inline]{Mencion a las lenguas de Arnold y el lockeo.}

\begin{figure*}[t]
    \centering
    \includegraphics[width=1\textwidth]{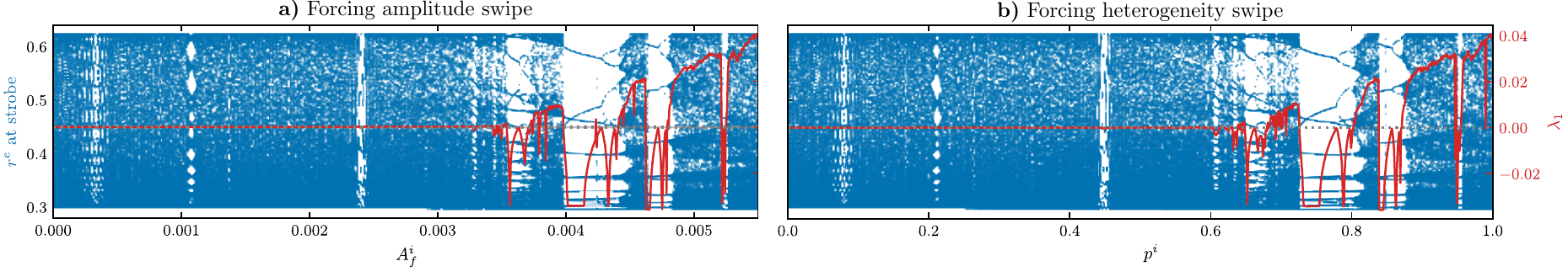}
    \caption{Stroboscopic plot of $r^e$ (blue) and corresponding Maximum Lyapunov Exponent $\lambda_1$ (red) of the forced system. Subfigure a) shows a sweep over the forcing amplitude applied to the entire inhibitory population, $A_f^i$. Subfigure b)) shows a sweep over the fraction of the inhibitory population for a fixed amplitude $A_f^i=0.0055$.}
    \label{fig:scan_A_scan_p}
\end{figure*}

To understand the unfolding of chaos from periodic orbits, we fix the frequency at $\Omega_f^i/\Omega=2.225$, and perform a sweep of increasing amplitude. After a transient, we record stroboscopic points of the coordinate $r^e$ as well as the Maximum Lyapunov Exponent of the system. This is shown in the left subplot of Figure~\ref{fig:scan_A_scan_p}.

% \todo[inline]{Ejemplo de barrido estroboscópico, exp. de Lyapunov, aparición de caos.}

% \begin{figure*}[t]
%     \centering
%     \includegraphics[width=1\textwidth]{Figures/strobe_MLE_Ascan.pdf}
%     \caption{Stroboscopic plot of $r^e$ (blue) and corresponding Maximum Lyapunov Exponent $\lambda_1$ (red) for each amplitude $A_f^i$ of the forcing, applied to the entire inhibitory population.}
%     \label{fig:scan_A}
% \end{figure*}

In general terms, we can distinguish three dynamical regimes: 
\begin{itemize}
    \item For small values of $A^i_f$ there is a dense area covered by the stroboscopic plot, corresponding to an invariant cycle $\mathbb{S}^1$, with angles coming from the $r^e$ coordinate, which is cyclic in the orbit. This is the projection of a quasiperiodic orbit on the torus $\mathbb{S}^1\times \mathbb{S}^1$, where the other angle is the time modulo the forcing period $T_f$. Since we are sampling points every forcing period, and that is not commensurable with the cycle's period $T$, even with $A^i_f\to0$ the entire torus is explored. 
    
    The associated MLE is zero (up to numerical error), since two nearby trajectories do not exponentially diverge nor converge.

    \item For some intermediate regions, the MLE drops to negative values, these correspond to the regions of period-$q$. There, the frequencies are locked, and the trajectories thus end up being together, exploring only a periodic curve on the full torus. Increasing the forcing the system undergoes a series of period-doubling bifurcations. 

    \item After each period-$q$ region, we have a somewhat dense one with positive MLE: this corresponds to chaos, and the torus having been broken.%: not only the solution is not periodic (like in the quasiperiodic case), but it is also no longer on the surface of the torus. It is slightly deformed, sometimes above sometimes below (via the usual mechanism of stretching and folding characteristic of chaotic attractors), creating a strange attractor.

\end{itemize}

% \todo[inline]{Mención al mecanismo de cuasiperiodicidad y ruptura de toros.}

\subsection{Heterogeneously forced populations at fixed amplitude}

We consider now the situation of an external forcing whose amplitude is set to a fixed value, although only affecting a fraction $p^i\in[0,1]$ of the inhibitory population. How does the transition from a periodic orbit to a chaotic regime occur, as a higher percentage of the population is affected by the forcing? % Is it a simple phase transition (first, second order) as more parts of the population are included in the forcing, or is it something else entirely? 

This setting is motivated by optogenetic experiments, where external stimulation typically affects only a subset of the targeted neurons. In particular, there have been experiments\cite{cardin2009driving, sohal2009parvalbumin} which showed that manipulating inhibitory neurons can strongly influence collective oscillations in cortical networks. Here, we take the fraction ($p^i$) of directly forced inhibitory neurons as a control parameter, allowing us to study how the extent of the perturbation affects the macroscopic dynamics.

We choose the same forcing frequency as in the previous subsection, $\Omega_f^i/\Omega=2.225$, and a value of the forcing amplitude of $A_f^i=0.0055$, so as to ensure the appearance of chaos at $p^i=1$ (the excitatory population remains unforced, $p^e=0$). It is thus clear that, as was the case with the homogeneous forcing sweep, we start from a regime with zero Lyapunov exponent and we move towards the same $\lambda_1$. It is unclear, however, how this transition takes place.    

% \todo[inline]{Referencia a las ecuaciones iniciales.}

\begin{figure*}[t]
    % \hspace{-0.4in}
    \includegraphics[width=1\linewidth]{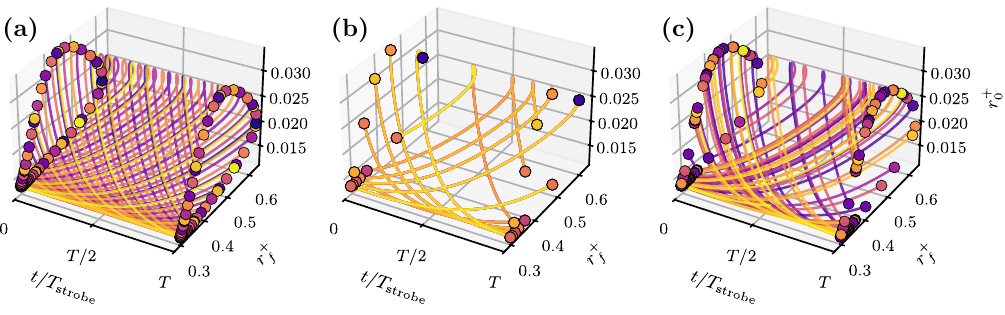}
    \caption{Trajectories obtained integrating 20 $T_f$ periods after an initial (discarded) transient, with the time compactified to $[0,T_f)$. The amplitude is fixed at $A_f^i=0.0055$ while the fractions of the population are $p^i=0.1$ (\textbf{a}, quasiperiodic), $p^i=0.75$ (\textbf{b}, period-11), $p^i=0.9$ (\textbf{c}, chaotic), respectively.}
    \label{fig:toribreak}
\end{figure*}

% \todo[inline]{Ejemplo de barrido estroboscópico, exp. de Lyapunov, aparición de caos.}
We present the result of the stroboscopic plot of the $r^e$ coordinate in the right subplots of Figure~\ref{fig:scan_A_scan_p}. 
% \begin{figure*}[t]
%     \centering
%     \includegraphics[width=1\linewidth]{Figures/strobe_MLE_pscan.pdf}
%     \caption{Stroboscopic plot of $r^e$ (blue) and corresponding Maximum Lyapunov Exponent $\lambda_1$ (red) for each fraction $p^i$ of the inhibitory population subject to forcing, with fixed amplitude $A_f^i=0.0055$. \textcolor{red}{Igual pasar esta figura a doble columnar juntada con la del otro escaneo estroboscópico}}
%     \label{fig:scan_p}
% \end{figure*}
As in the previous exploration, we find three regimes: quasiperiodicity, periodic motion with period doubling bifurcations, and lastly, chaos. An example of each of the three regimes is shown in Figure~\ref{fig:toribreak}, where a trajectory is integrated and represented over a interval of length $T_f$ (after a transient), where $T_f=2\pi/\Omega_f^i$. The toroidal structure (upon identification $0\sim T_f$) is clearly seen underlying the system's dynamics, which is broken at the onset of chaos.

%{\color{red}
%\paragraph{Microscopic check}

%It must be noted, however, that these calculations were all based on arguments which are strictly valid in the thermodynamic limit; in finite systems fractions of the population create an imbalance which is not possible to compensate in the homogeneous forcing case. }

%\todo[inline]{Y si miramos esto??? mirar el efecto de tener un conjunto de neuronas forzadas (mirar en p= 0.5 vs p = 0.9). Hacer raster plot de inhibidores forzados y por otro lado los no forzados. Calcular mediante el método de Wolf el exponente de la serie temporal de la media en ambos casos y comprobar que sea caótico y periódico en cada caso.}
% {\color{purple}

\begin{figure}[t]
\centering
\includegraphics[width=\linewidth]{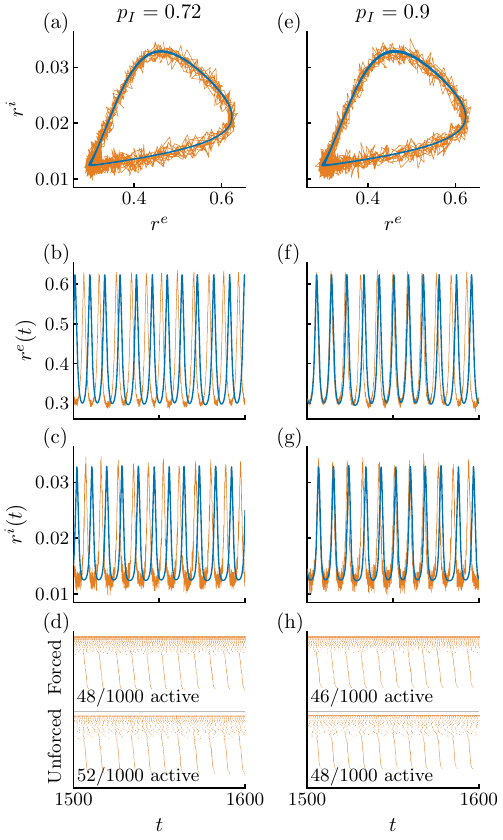}
\caption{Mean-field (blue) and microscopic (orange) responses
under heterogeneous inhibitory forcing, for $p_I=0.72$ (left)
and $p_I=0.90$ (right), with $A_f^i=0.0055$ and
$\Omega_f^i/\Omega_0=2.225$.
(a,e) Firing-rate projections; (b,f) excitatory and (c,g)
inhibitory firing rates. (d,h) Inhibitory spike rasters for
1000 sampled forced neurons (upper block) and 1000 unforced
neurons (lower block), ordered by increasing excitability.
Only the active rank range is displayed; all sampled spikes
in the shown interval are retained.}
\label{fig:forced_micro}
\end{figure}

The mean-field response is periodic at $p_I=0.72$ and chaotic
at $p_I=0.90$. Figure~\ref{fig:forced_micro} compares these
responses with finite-network simulations containing
$N_e=N_i=262\,144$ neurons. Both cases use
$(\eta_0^e,\eta_0^i)=(-1.419,-3.38)$,
$\Delta_e=\Delta_i=0.1$, $\tau_e=\tau_i=0.01$, and
$(J_{ee},J_{ei},J_{ie},J_{ii})=(8,6,6,4)$,
with negative inhibitory inputs. The forced inhibitory neurons
receive $I_f^i(t)=0.0055\cos(\Omega_f^i t)$, where
$\Omega_f^i=2.225(2\pi/9.112098)$.

Excitabilities are assigned deterministically as
$\eta_j=\eta_0+\Delta\tan[\pi(j/(N_g+1)-1/2)]$,
$j=1,\ldots,N_g$, separately within the excitatory population
and each inhibitory subgroup of size $N_g$.
The integration step is $0.0025$. Microscopic population rates
include all neurons and use time bins of width $0.05$.
The interval $1500\leq t\leq1600$ is shown without temporal
realignment; raster subsampling affects visualization only.
% }

\begin{figure*}[t]
    \centering
    \includegraphics[width=\textwidth]{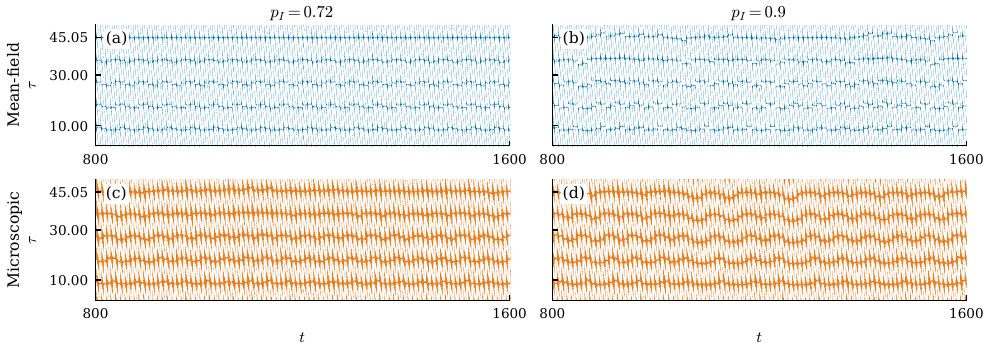}
    \caption{
        Close-return diagrams of the inhibitory firing rate
        for $p_I=0.72$ (left) and $p_I=0.90$ (right), from
        the mean-field model (a,b, blue) and microscopic
        network (c,d, orange), using the simulations compared
        in the preceding time-series figure.
        Colored points indicate
        $|r^i(t+\tau)-r^i(t)|<0.005\,\Delta r^i$,
        where $\Delta r^i$ is the range of each signal over
        $800\leq t<1700$.
        The full comparison interval $800\leq t<1600$
        is shown for delays $2\leq\tau\leq50$.
    }
    \label{fig:close_returns_inhibitory}
\end{figure*}

To complement the time-series comparison,
Fig.~\ref{fig:close_returns_inhibitory} shows scalar close-return
diagrams of the inhibitory firing rates, following
Mindlin et al.~\cite{mindlin1991topological}.
The periodic mean-field response at $p_I=0.72$ exhibits
persistent horizontal returns near $\tau=45.05$, whereas
the chaotic trajectory at $p_I=0.90$ exhibits local return
segments without the corresponding global repetition,
consistent with the Lyapunov analysis.
The microscopic population rates show more regular recurrence
bands at $p_I=0.72$ and stronger temporal modulation at
$p_I=0.90$, in qualitative agreement with the contrasting
temporal organization of the corresponding mean-field responses.

What is most intriguing at this point is the clear resemblance between the left and right subplots of Figure~\ref{fig:scan_A_scan_p}. Not only do they display the same dynamical regimes, but there is a great qualitative agreement on both the stroboscopic plot as well as the computed MLEs. This is by no means expected: the homogeneous forcing part concerns an eight-dimensional, non-autonomous dynamical system, whereas this heterogeneously forced system is ten-dimensional\footnote{The original forced system we derived in \eqref{eq:QIFmacroscopicForced} is twelve-dimensional, however if the excitatory population is not forced their collective variables do not double their dimensionality.}. 

This suggests that there is a mapping between the amplitude of the forcing which the entire ensemble of excitable units is subject to, and the fraction of population in the ensemble which is subject to a given forcing. In other words, if one forces a neural population with a percentage $p$ of the maximum amplitude of the forcing, the result is almost the same as if one had forced a percentage $p$ of the population at maximum amplitude. This is indeed the case, as we show in the next subsection.

% \textcolor{purple}{Quizás para el título a la sección y subsecciones pondría; IV Dynamics under heterogeneous forcing: Two routes to collective chaos y a priori el IV.A Varying forcing amplitude over the entire population y el IV.B Heterogeneously forced populations at fixed amplitude; me parecen bien y no se si añadiría un IV.C Comparing the two routes.
% En general veo que los resultados micro van a ir con los macro en todo momento, que no van a tener una sección completa me refiero. Me parece bien, pero otra opción sería hacer una sección solo micro (a modo de idea).}%

% \textcolor{red}{Lo de two routes to collective chaos yo no lo pondría, porque las rutas al caos son otra cosa. Esto es parámetros distintos que llevan, por la misma ruta, al caos}

\subsection{Equivalence between amplitude and forced fraction}

% \todo[inline]{Observación de que obtenemos lo mismo de ambas maneras.}
In the previous subsection we noted the striking similarity between the two parameter sweeps. Figure~\ref{fig:scan_A_scan_p} suggested that the effect of the forcing over a fraction of the population was smeared out in such a way as to make it equivalent to a stronger forcing over a smaller fraction, and vice versa.  

\begin{figure}[h]
    \centering
    \includegraphics[width=0.95\linewidth]{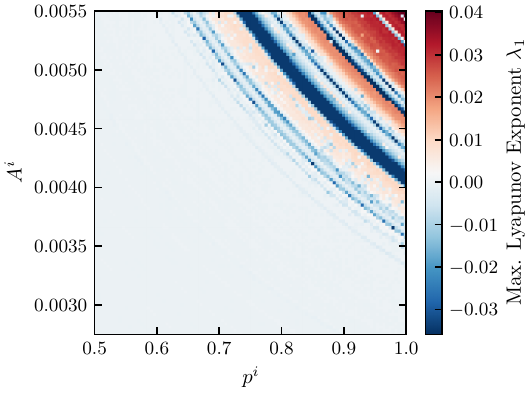}
    \caption{Heatmap of the Maximum Lyapunov Exponent $\lambda_1$ for each pair $(p^i,A_f^i)$.}
    \label{fig:doublescan_Ap}
\end{figure}

We provide further numerical evidence of this phenomenon by computing the MLE of \eqref{eq:QIFmacroscopicForced} for different values of fraction $p^i\in[0.5,1]$ and amplitude $A_f^i\in(0.003,0.0055)$, as shown in Figure~\ref{fig:doublescan_Ap}, again with fixed frequency $\Omega_f^i/\Omega=2.225$. This figure turns out to be nearly symmetric with respect to the off-diagonal, signaling an approximate mapping between the two parameters involved in the forcing.

It turns out that this can be understood analytically, through a rewriting of the heterogeneously forced system's variables which brings it to a homogeneous form. For simplicity, let us focus on the inhibitory variables, since said population is the one split by the heterogeneous forcing. Define the population means as $\bar{r}^i=p^i r^i_f+(1-p^i)r_u^i$ and $\bar{v}^i=p^i v^i_f+(1-p^i)v_u^i$. Note that the following identity holds:
\begin{equation}
    p\, a_f b_f + (1-p)\,a_u b_u= \bar{a}\bar{b} + p(1-p)(a_f-a_u)(b_f-b_u).
\end{equation}

Averaging the subpopulation equations with weights $p^i$ and $(1-p^i)$ then gives, with no approximation,
\begin{align}
    \displaystyle \dv{\bar{r}^i}{t} &= \dfrac{\Delta^i}{\pi} + 2\bar{v}^i \bar{r}^i + 2p^i(1-p^i)(r^i_f-r^i_u)(v^i_f-v^i_u), \nonumber \\[0.1in]
    \displaystyle \dv{\bar{v}^i}{t} &= \eta_0^i + p^i A_f^i\cos(\Omega^i_f t) + J^{ie} s^e - J^{ii}s^i - \pi^2 (\bar{r}^i)^2 + (\bar{v}^i)^2 \nonumber \\[0.1in]
    &\quad + p^i(1-p^i)\left[(v^i_f-v^i_u)^2-\pi^2(r^i_f-r^i_u)^2\right].
\end{align}

Note that this has effectively reduced the dynamics of the entire inhibitory population to a homogeneously forced one with amplitude $p^i A_f^i$, plus corrections which are of order $\mathcal{O}(p^i(1-p^i)(A_f^i)^2)$. These corrections are due to the mismatch between forced and unforced variables, which are driven only by the forcing $\mathcal{O}(A_f^i)$. This is supported by Floquet analysis of the unforced cycle, with multipliers corresponding to $r^i,v^i$ of order $10^{-8}$. Figure~\ref{fig:forcing_scan} is thus understood as displaying approximately the same exponent along hyperbolic level curves $p^i\propto 1/A_f^i$. %The smallness of $A_f^i$ is why the hyperbolic curves resemble straight lines.   

Even though we derived this equivalence for the inhibitory variables, the whole network inherits the equivalence since $\tau^i\dot w^i = - w^i + \bar{r}^i$ is exactly the homogeneous form. This is all in accordance with Figures~\ref{fig:scan_A_scan_p} and \ref{fig:doublescan_Ap}: the correction vanishes at $p^i=0$ and $p^i=1$, and is largest at $p^i=0.5$ which is why deviations start appearing around that point.

%%%%%%%%%%%%%%%%%%%%%%%%%%%%%%%%%%%%%%%%

% \newpage
%%%%%%%%%%%%%%%%%%%%%%%%%%%%%%%%%%%%%%%%
\section{\label{sec:conclusions}Discussion and Conclusions}

% \section{Conclusions}

Over the last decade, exact mean-field reductions have made it possible to
derive low-dimensional descriptions of neural populations directly from the
dynamics of spiking networks~\cite{ott2008low,luke2013complete,montbrio2015macroscopic,bick2020understanding},
providing a microscopic foundation for models that were previously postulated
phenomenologically~\cite{wilson1972excitatory}. A growing body of literature
explores the dynamical richness that these models acquire once biologically
motivated ingredients are included, such as synaptic
kinetics~\cite{devalle2017firing,coombes2016next} or periodic
inputs~\cite{schmidt2018network,ceni2020cross}. In this work
we have combined three of these ingredients, namely interacting excitatory
and inhibitory populations, a double synaptic filter and periodic forcing,
and characterized the collective dynamics that arise from them.

For the unforced dynamics, numerical continuation together with network
simulations reveals a rich landscape as the mean excitabilities of both
populations are varied: nine dynamical regimes, featuring collective
oscillations, bistability between quiescence and oscillations, and
hysteresis. Part of this structure, such as the cusp and the pair of Hopf
branches, is shared with the Wilson--Cowan
model~\cite{hoppensteadt1997weakly,roulet2016average}, but several features
are not, notably a stable limit cycle born at a homoclinic bifurcation, which
coexists with a quiescent state; its stability is guaranteed by the negative
saddle quantity of the intermediate equilibrium. These features stem from the
interplay between firing rate and mean membrane potential that the exact
reduction retains~\cite{montbrio2015macroscopic,devalle2017firing}. For the
fast synapses considered here, the double filter leaves the fixed points and
the saddle-node bifurcations unchanged and only shifts the Hopf boundaries.
Despite the eight-dimensional phase space, we found no evidence of chaos in
the autonomous system. Floquet analysis indicates why, at least for the
oscillatory attractors we examined: the limit cycle lies on a strongly
attracting two-dimensional invariant manifold, on which chaotic dynamics are
excluded. We leave for future work the regime in which synaptic and membrane
time scales are comparable, which is also the biologically realistic one and
where the double filter is expected to play a more prominent role.

When the inhibitory population is periodically forced close to the
homoclinic bifurcation, even weak forcing elicits a non-trivial response:
Arnold tongues, in which the collective rhythm locks to the forcing at
rational frequency ratios, quasiperiodic motion between them, and chaotic
regions with positive largest Lyapunov exponent. The observed sequence of
locking, period doubling within the tongues and breakup of the invariant
torus is consistent with the torus-breakdown
scenario. The proximity of the homoclinic orbit also makes the rhythm fragile: since the cycle passes close to a saddle, moderate forcing amplitudes suffice to push the system onto the coexisting quiescent state. For a membrane time constant of $10$~ms, the collective rhythm lies in the alpha band ($\approx 11$~Hz) and the forcing in the beta range ($\approx 24$~Hz), a range in which periodic drive has been shown to destabilize persistent network
states~\cite{schmidt2018network}. Our results thus complement previous
studies of forced next-generation models, in which the forced rhythm was born
at a Hopf bifurcation~\cite{ceni2020cross}.

Moreover, forcing only a fraction $p^i$ of the inhibitory neurons with
amplitude $A_f^i$ produces nearly the same dynamics, including the Lyapunov
exponents, as forcing the whole population with amplitude $p^iA_f^i$. This is
far from obvious a priori, since the two situations correspond to dynamical
systems of different dimension: splitting the population doubles its degrees
of freedom. We showed that an exact rewriting in terms of population means
turns the heterogeneously forced system into a homogeneously forced one with
effective amplitude $p^iA_f^i$, plus corrections of order
$p^i(1-p^i)(A_f^i)^2$ arising from the mismatch between forced and unforced
neurons. This mismatch contracts at a rate set by the heterogeneity of the
excitabilities, which therefore underlies the equivalence.

This amplitude--fraction  depends neither on the synaptic kinetics nor on the specific dynamical regime: it only requires statistically identical subpopulations that interact through population averages, and a contracting mismatch between them. We have shown that the equivalence holds well beyond synchronization thresholds, across locking, torus breakdown and chaos, and we have quantified its corrections. For neural populations, it implies that the collective response to a rhythmic input is set, to leading order, by the product of its strength and the fraction of neurons it reaches: driving few neurons strongly or many neurons weakly is nearly equivalent, as long as the forced fraction is not too small. This suggests a trade-off between stimulus intensity and the fraction of responsive cells.

Our analysis relies on the thermodynamic limit, Lorentzian heterogeneity and
all-to-all, current-based coupling. Finite-size effects are expected to
matter most for small forced fractions, where the forced subpopulation
contains few neurons. The weaker contraction of the excitatory
mismatch also predicts a less accurate equivalence when a fraction of the
excitatory population is forced. Finally, extending the analysis to
comparable synaptic and membrane time scales, conductance-based synapses and
sparse connectivity would clarify how far these results carry over to more
realistic networks.

% \todo[inline]{We hypothesize that this might also happen for more general ensembles of dynamical systems. This is, however, beyond th escope of this manuscript.}

%%%%%%%%%%%%%%%%%%%%%%%%%%%%%%%%%%%%%%%%

%%%%%%%%%%%%%%%%%%%%%%%%%%%%%%%%%%%%%%%%
% \subsection*{Code and data availability.}

\section*{Acknowledgements.}

G. C-A. acknowledges funding from project 2025/00767/001 (UC3M). F. J. M-R. acknowledges funding from a predoctoral contract from Universidad Rey Juan Carlos (Contratos predoctorales propios de la URJC para Personal Investigador en Formación 2025). G. B. M. acknowledges partial support from Universidad de Buenos Aires and CONICET..

\appendix

%%%%%%%%%%%%%%%%%%%%%%%%%%%%%%%%%%%%%%%%
\section{The mean-field equation derivation}\label{appen:OA}

We briefly show the derivation of equations \eqref{eq:QIFmacroscopic}. We consider an ensemble of all-to-all coupled theta neurons as in \eqref{eq:thetamodel_excitinhib}, under no external influence. The influence of forcing on respective subpopulations is discussed in section~\ref{sec:forced} as an extension of this derivation.

We consider the thermodynamic limit of \eqref{eq:thetamodel_excitinhib}, in which the system is described by the density of neurons with excitability $\eta^a$ at phase $\theta$ at time $t$, $f(\theta,\eta^a,t)$. This density satisfies
\begin{equation}
    \int_0^{2\pi}f(\theta,\eta^a, t)\, \text{d}\theta = g(\eta^a).
\end{equation}
with $g(\eta^a)$ a Lorentzian distribution \eqref{eq:lorentzian}. We can expand the density in Fourier modes, and impose the Ott-Antonsen ansatz \cite{ott2008low} $\alpha_n(\eta,t)=\alpha(\eta,t)^n$, upon which we have
\begin{equation}
f(\theta,\eta^a,t)
= \frac{g(\eta^a)}{2\pi}
\left[1 + \sum_{n=1}^{\infty} \left( \alpha(\eta,t)^n e^{in\theta} 
+ \text{c.c.} \right)\right],
\end{equation}
where c.c. stands for the complex conjugate.

Inserting this expansion into the continuity equation
$\partial_t f^a + \partial_\theta(\dot\theta f^a)=0$, all Fourier modes are satisfied provided
\begin{equation}
    \partial_t\alpha = -i(1+I^a)\,\alpha + \frac{i}{2}(1-I^a)(1+\alpha^2).
\end{equation}
Assuming that $\bar\alpha(\eta,t)$ can be analytically continued into the
upper half of the complex $\eta$-plane, with $|\alpha|\le1$ there
\cite{ott2008low}, the integral over $\eta$ can be closed in that half-plane
and only the pole of $g^a$ at $\eta_0^a+i\Delta^a$ contributes. Hence
\begin{equation}
    z^a(t)=\int_{-\infty}^{\infty} g^a(\eta)\,\bar\alpha(\eta,t)\,\text{d}\eta
          =\bar\alpha(\eta_0^a+i\Delta^a,t),
\end{equation}
and evaluating the complex conjugate of the equation for $\alpha$ at
$\eta=\eta_0^a+i\Delta^a$ yields
\begin{equation}
    \dv{z^a}{t} = \displaystyle 
        - \dfrac{i}{2}(1-z^a)^{2} +\dfrac{1}{2}(1+z^a)^2[i(\eta_0^a+J^{ae} s^e - J^{ai} s^i) - \Delta^a].
\end{equation}

Defining $\zeta^a\equiv(1-z^a)/(1+z^a)$, the equation for $z^a$ becomes
\begin{equation}
    \dv{\zeta^a}{t} = \Delta^a - i\left(\eta_0^a+J^{ae}s^e-J^{ai}s^i\right)
                      + i(\zeta^a)^2 .
\end{equation}
Since $\mathrm{Re}\,\zeta^a=(1-|z^a|^2)/|1+z^a|^2=\pi\phi^a$, the real part of
$\zeta^a$ is proportional to the firing rate, while its imaginary part equals
minus the (principal-value) mean membrane potential \cite{montbrio2015macroscopic}.
Writing $\zeta^a=\pi r^a - i v^a$ and separating real and imaginary parts
gives the first two equations of \eqref{eq:QIFmacroscopic}.

Regarding the synaptic variables, averaging the equations for $s_n^a$ and
$w_n^a$ over each population gives 
\begin{equation}
    \tau^a\dot s^a=-s^a+w^a,\quad 
    \tau^a\dot w^a=-w^a+\frac{1}{N^a}\sum_n\sum_k\delta(t-t_n^k),
\end{equation}
where the last term converges to the population firing rate $\phi^a(t)=r^a(t)$ in the thermodynamic limit. Since $s^e$ and $s^i$ enter the dynamics of every neuron as a common, time-dependent input, independent of $\theta$ and $\eta$, they only make $I^a$
time dependent: the Ott--Antonsen ansatz and the residue evaluation above are
unaffected, and the reduction remains exact with the synaptic cascade.

It is worth mentioning that restricting the dynamics to the Ott-Antonsen manifold is justified for heterogeneous populations\cite{ott2009long}, whereas for parameter-dependent systems such as theta and QIF neurons the attractiveness has also been shown\cite{pietras2016ott}.

\section{Adimensionalization of the model}\label{appen:adim}

We can obtain a minimal description of the model \eqref{eq:QIFmacroscopic} by adimensionalization. For that, We introduce the rescaled variables
\begin{equation}
    t = t_0 T, \quad (r^a, v^a, s^a, w^a) = \frac{1}{t_0}\,(R^a, V^a, S^a, W^a),
    \quad t_0 = \sqrt{\frac{\pi}{\Delta^e}},
\end{equation}
with $a\in\{e,i\}$. Since both populations share the same time scale, only one
of the widths can be absorbed; we choose $\Delta^e$. The system becomes
\begin{align}
    \dv{R^a}{T} &= \tilde{\Delta}^a + 2V^a R^a,  \nonumber \\
    \dv{V^a}{T} &= \tilde{\eta}_0^a + \tilde{J}^{ae}S^e - \tilde{J}^{ai}S^i
                   - \pi^2 (R^a)^2 + (V^a)^2, \nonumber\\
    \tilde{\tau}^a \dv{S^a}{T} &= -S^a + W^a, \qquad
    \tilde{\tau}^a \dv{W^a}{T} = -W^a + R^a,
\end{align}
with rescaled parameters
\begin{equation}
    \tilde{\Delta}^a = \frac{\Delta^a}{\Delta^e}, \qquad
    \tilde{\eta}_0^a = \frac{\pi}{\Delta^e}\,\eta_0^a, \qquad
    \tilde{J}^{ab} = t_0 J^{ab}, \qquad
    \tilde{\tau}^a = \frac{\tau^a}{t_0}.
\end{equation}
Hence $\tilde{\Delta}^e = 1$, while the ratio $\tilde{\Delta}^i = \Delta^i/\Delta^e$
remains a free parameter.

\section{Analytical expressions for the Jacobian and saddle-node bifurcations}\label{appen:calc}

Here we examine in more detail the unforced model described by equation \eqref{eq:QIFmacroscopic}, whose fixed points equations were shown in \eqref{eq:fixedpointsQIFexcitinhib}. These fixed point equations involve a nontrivial, polynomial relation between the firing rates of the excitatory and inhibitory populations, $r^e$ and $r^i$ respectively, with the rest of the dynamical variables algebraically related to them.

The Jacobian of the system is
\begin{widetext}\begin{equation}
    D_F(\mathbf{x})= \left(
    \begin{array}{cccc|cccc}
            2v^e & 2r^e & 0 & 0               & 0 & 0 & 0 & 0\\
            -2\pi^2 r^e & 2v^e & J^{ee} & 0   & 0 & 0 & -J^{ei} & 0\\
            0 & 0 & -1/\tau^e & 1/\tau^e      & 0 & 0 & 0 & 0\\
            1/\tau^e & 0 & 0 & -1/\tau^e      & 0 & 0 & 0 & 0\\
            \hline
            0 & 0 & 0 & 0                     & 2v^i & 2r^i & 0 & 0 \\
            0 & 0 & J^{ie} & 0                & -2\pi^2 r^i & 2v^i & -J^{ii} & 0 \\
            0 & 0 & 0 & 0                     & 0 & 0 & -1/\tau^i & 1/\tau^i\\
            0 & 0 & 0 & 0                     & 1/\tau^i & 0 & 0 & -1/\tau^i     
        \end{array}\right).
    \end{equation}
\end{widetext}

We can see, as we advanced in the main text, that the Jacobian does not depend on $\Delta^e$ or $\Delta^i$. Moreover, their value does not affect the existence of fixed points, see \eqref{eq:fixedpointsQIFexcitinhib}. Both of these are related to the fact that these parameters can be removed by adimensionalization, as shown in Appendix~\ref{appen:adim}: their values do not affect the existence or stability of fixed points.

The same can not be said about $\eta_0^e,\, \eta_0^i$, which do affect the existence of fixed points, as we can clearly see in the phase diagram of Figure~\ref{fig:bifdiagram}. The amount of fixed points does change when we move across the saddle-node bifurcation curves. Beyond the usage of numerical continuation methods, those curves can be analytically characterized.

We parametrize the saddle-node curves by $r^e=\lambda$. The saddle-node condition $\det(D_F(\mathbf{x}))=0$ can be rewritten as
\begin{align}
    &2\pi^2 (r^i)^4 + \frac{(\Delta^i)^2}{2\pi^2} \nonumber \\
     &+ \left(J^{ii} + \frac{2\pi^2 J^{ei}J^{ie} \lambda^3}{4\pi^4 \lambda^4 - 2\pi^2 J^{ee} \lambda^3 + (\Delta^+)^2}\right)(r^i)^3 = 0,
\end{align}
where $\lambda>0$ ranges over the interval in which $4\pi^4 \lambda^4 - 2\pi^2 J^{ee} \lambda^3 + (\Delta^e)^2 < 0$. The saddle-node bifurcation curves in the $(\eta_0^e,\eta_0^i)$ plane are then given parametrically by \eqref{eq:SNparametric} with one branch of the curve for each positive root $r^i(\lambda)$ of the quartic.

% \section{\color{red} Notes on numerical simulations}\label{appen:numerics}

% \nocite{*}
\bibliography{bibliography}% Produces the bibliography via BibTeX.

@article{rall1967distinguishing,
  title={Distinguishing theoretical synaptic potentials computed for different soma-dendritic distributions of synaptic input.},
  author={Rall, Wilfrid},
  journal={Journal of Neurophysiology},
  volume={30},
  number={5},
  pages={1138--1168},
  year={1967}
}

@article{mindlin1991topological,
  title={Topological analysis of chaotic time series data from the Belousov-Zhabotinskii reaction},
  author={Mindlin, G. B. and Solari, H. G. and Natiello, M. A. and Gilmore, R. and Hou, X.-J.},
  journal={Journal of Nonlinear Science},
  volume={1},
  number={2},
  pages={147--173},
  year={1991},
  publisher={Springer}
}

@article{wilson1972excitatory,
  title={Excitatory and inhibitory interactions in localized populations of model neurons},
  author={Wilson, Hugh R and Cowan, Jack D},
  journal={Biophysical journal},
  volume={12},
  number={1},
  pages={1--24},
  year={1972},
  publisher={Elsevier}
}

@article{kopell1986subcellular,
  title={Subcellular oscillations and bursting},
  author={Kopell, N and Ermentrout, GB},
  journal={Mathematical Biosciences},
  volume={78},
  number={2},
  pages={265--291},
  year={1986},
  publisher={Elsevier}
}

@book{hoppensteadt1997weakly,
  title={Weakly connected neural networks},
  author={Frank C. Hoppensteadt and Eugene M. Izhikevich},
  year={1997},
  isbn = {978-0-387-94948-2},
  publisher = {Springer},
  series = {Applied Mathematical Sciences},
  edition = 1
}

@book{Kuznetsov1998,
  address = {Berlin},
  author = {Kuznetsov, Yuri A.},
  edition = 2,
  isbn = {978-0387983820},
  number = 112,
  publisher = {Springer},
  series = {Applied Mathematical Sciences},
  title = {Elements of applied bifurcation theory},
  year = 1998
}

@book{wilson1999spikes,
  title={Spikes, decisions, and actions: the dynamical foundations of neuroscience},
  author={Wilson, Hugh Reid},
  publisher={Oxford University Press},
  year={1999}
}

@article{buzsaki2002theta,
  title={Theta oscillations in the hippocampus},
  author={Buzs{\'a}ki, Gy{\"o}rgy},
  journal={Neuron},
  volume={33},
  number={3},
  pages={325--340},
  year={2002},
  publisher={Elsevier}
}

@article{borgers2003synchronization,
  title={Synchronization in networks of excitatory and inhibitory neurons with sparse, random connectivity},
  author={B{\"o}rgers, Christoph and Kopell, Nancy},
  journal={Neural computation},
  volume={15},
  number={3},
  pages={509--538},
  year={2003},
  publisher={MIT Press}
}

@book{wiggins2003introduction,
  title={Introduction to applied nonlinear dynamical systems and chaos},
  author={Wiggins, Stephen},
  year={2003},
  publisher={Springer}
}

@book{izhikevich2007dynamical,
  title={Dynamical systems in neuroscience},
  author={Izhikevich, Eugene M},
  year={2007},
  publisher={MIT press}
}

@article{ott2008low,
  title={Low dimensional behavior of large systems of globally coupled oscillators},
  author={Ott, Edward and Antonsen, Thomas M},
  journal={Chaos: An Interdisciplinary Journal of Nonlinear Science},
  volume={18},
  number={3},
  year={2008},
  publisher={AIP Publishing}
}

@article{ott2009long,
  title={Long time evolution of phase oscillator systems},
  author={Ott, Edward and Antonsen, Thomas M},
  journal={Chaos: An interdisciplinary journal of nonlinear science},
  volume={19},
  number={2},
  year={2009},
  publisher={AIP Publishing}
}

@article{sohal2009parvalbumin,
  title={Parvalbumin neurons and gamma rhythms enhance cortical circuit performance},
  author={Sohal, Vikaas S and Zhang, Feng and Yizhar, Ofer and Deisseroth, Karl},
  journal={Nature},
  volume={459},
  number={7247},
  pages={698--702},
  year={2009},
  publisher={Nature Publishing Group UK, London}
}

@article{cardin2009driving,
  title={Driving fast-spiking cells induces gamma rhythm and controls sensory responses},
  author={Cardin, Jessica A and Carl{\'e}n, Marie and Meletis, Konstantinos and Knoblich, Ulf and Zhang, Feng and Deisseroth, Karl and Tsai, Li-Huei and Moore, Christopher I},
  journal={Nature},
  volume={459},
  number={7247},
  pages={663--667},
  year={2009},
  publisher={Nature Publishing Group UK, London}
}

@article{olmi2010collective,
  title={Collective chaos in pulse-coupled neural networks},
  author={Olmi, Simona and Politi, Antonio and Torcini, Alessandro},
  journal={EPL (Europhysics Letters)},
  volume={92},
  number={6},
  pages={60007},
  year={2010}
}

@article{luke2013complete,
  title={Complete classification of the macroscopic behavior of a heterogeneous network of theta neurons},
  author={Luke, Tanushree B and Barreto, Ernest and So, Paul},
  journal={Neural computation},
  volume={25},
  number={12},
  pages={3207--3234},
  year={2013},
  publisher={MIT Press One Rogers Street, Cambridge, MA 02142-1209, USA journals-info~…}
}

@article{laing2014derivation,
  title={Derivation of a neural field model from a network of theta neurons},
  author={Laing, Carlo R},
  journal={Physical Review E},
  volume={90},
  number={1},
  pages={010901},
  year={2014},
  publisher={APS}
}

@article{so2014networks,
  title={Networks of theta neurons with time-varying excitability: Macroscopic chaos, multistability, and final-state uncertainty},
  author={So, Paul and Luke, Tanushree B and Barreto, Ernest},
  journal={Physica D: Nonlinear Phenomena},
  volume={267},
  pages={16--26},
  year={2014},
  publisher={Elsevier}
}

@article{montbrio2015macroscopic,
  title={Macroscopic description for networks of spiking neurons},
  author={Montbri{\'o}, Ernest and Paz{\'o}, Diego and Roxin, Alex},
  journal={Physical Review X},
  volume={5},
  number={2},
  pages={021028},
  year={2015},
  publisher={APS}
}

@article{roulet2016average,
  title={Average activity of excitatory and inhibitory neural populations},
  author={Roulet, Javier and Mindlin, Gabriel B},
  journal={Chaos: An Interdisciplinary Journal of Nonlinear Science},
  volume={26},
  number={9},
  year={2016},
  publisher={AIP Publishing}
}

@article{pazo2016quasiperiodic,
  title = {From Quasiperiodic Partial Synchronization to Collective Chaos in Populations of Inhibitory Neurons with Delay},
  author = {Paz\'o, Diego and Montbri\'o, Ernest},
  journal = {Physical Review Letters},
  volume = {116},
  issue = {23},
  pages = {238101},
  numpages = {5},
  year = {2016},
  month = {Jun},
  publisher = {American Physical Society},
}

@article{pietras2016ott,
    author = {Pietras, Bastian and Daffertshofer, Andreas},
    title = {Ott-{Antonsen} attractiveness for parameter-dependent oscillatory systems},
    journal = {Chaos: An Interdisciplinary Journal of Nonlinear Science},
    volume = {26},
    number = {10},
    pages = {103101},
    year = {2016},
    month = {10},
    issn = {1054-1500}
}

@article{coombes2016next,
  title={Next generation neural mass models},
  author={Coombes, Stephen and Byrne, Aine},
  journal={arXiv preprint arXiv:1607.06251},
  year={2016}
}

@article{devalle2017firing,
  title={Firing rate equations require a spike synchrony mechanism to correctly describe fast oscillations in inhibitory networks},
  author={Devalle, Federico and Roxin, Alex and Montbri{\'o}, Ernest},
  journal={PLoS computational biology},
  volume={13},
  number={12},
  pages={e1005881},
  year={2017},
  publisher={Public Library of Science San Francisco, CA USA}
}

@article{schmidt2018network,
  title={Network mechanisms underlying the role of oscillations in cognitive tasks},
  author={Schmidt, Helmut and Avitabile, Daniele and Montbri{\'o}, Ernest and Roxin, Alex},
  journal={PLoS computational biology},
  volume={14},
  number={9},
  pages={e1006430},
  year={2018},
  publisher={Public Library of Science, San Francisco, CA USA}
}

@misc{uribarri2019resonant,
      title={Resonant features in a forced population of excitatory neurons}, 
      author={Gonzalo Uribarri and Gabriel B. Mindlin},
      year={2019},
      eprint={1902.06008},
      archivePrefix={arXiv},
      primaryClass={nlin.AO},
}

@article{boari2019observable,
  title={Observable for a large system of globally coupled excitable units},
  author={Boari, Santiago and Uribarri, Gonzalo and Amador, Ana and Mindlin, Gabriel B},
  journal={Mathematical and Computational Applications},
  volume={24},
  number={2},
  pages={37},
  year={2019},
  publisher={MDPI}
}

@article{dumont2019macroscopic,
  title={Macroscopic phase resetting-curves determine oscillatory coherence and signal transfer in inter-coupled neural circuits},
  author={Dumont, Gr{\'e}gory and Gutkin, Boris},
  journal={PLoS computational biology},
  volume={15},
  number={5},
  pages={e1007019},
  year={2019},
  publisher={Public Library of Science San Francisco, CA USA}
}

@article{bick2020understanding,
  title={Understanding the dynamics of biological and neural oscillator networks through exact mean-field reductions: a review},
  author={Bick, Christian and Goodfellow, Marc and Laing, Carlo R and Martens, Erik A},
  journal={The Journal of Mathematical Neuroscience},
  volume={10},
  number={1},
  pages={9},
  year={2020},
  publisher={Springer}
}

@article{ceni2020cross,
  title={Cross frequency coupling in next generation inhibitory neural mass models},
  author={Ceni, Andrea and Olmi, Simona and Torcini, Alessandro and Angulo-Garcia, David},
  journal={Chaos: An Interdisciplinary Journal of Nonlinear Science},
  volume={30},
  number={5},
  year={2020},
  publisher={AIP Publishing}
}

@article{barrio2024synaptic,
  title = {Synaptic dependence of dynamic regimes when coupling neural populations},
  author = {Barrio, Roberto and Jover-Galtier, Jorge A. and Mayora-Cebollero, Ana and Mayora-Cebollero, Carmen and Serrano, Sergio},
  journal = {Physical Review E},
  volume = {109},
  issue = {1},
  pages = {014301},
  numpages = {9},
  year = {2024},
  month = {Jan},
  publisher = {American Physical Society},
}

@article{mayora2025dynamics,
  title={Dynamics of coupled neural populations: The role of synaptic dynamics},
  author={Mayora-Cebollero, Ana and Barrio, Roberto and Li, Lei and Mayora-Cebollero, Carmen and P{\'e}rez, Luc{\'\i}a},
  journal={Chaos: An Interdisciplinary Journal of Nonlinear Science},
  volume={35},
  number={6},
  year={2025},
  publisher={AIP Publishing}
}

% \textcolor{red}{Cuidado,  hay que homogeneizarlas!!!!!!!}

\end{document}